\documentclass[%
reprint,
 amsmath,amssymb,
 aps,
prb,
]{revtex4-2}

\usepackage{graphicx}% Include figure files
\usepackage{dcolumn}% Align table columns on decimal point
\usepackage{bm}% bold math
\usepackage{mathrsfs}
\usepackage{braket}
\usepackage{color}
\usepackage{comment}
\usepackage{hyperref}% add hypertext capabilities
\hypersetup{
     colorlinks   = true,
     linkcolor    = blue,
     citecolor    = blue,
     urlcolor     = blue
}
\newcommand{\q}{\bm{q}}
\newcommand{\J}{\bm{J}}
\newcommand{\R}{\mathrm{Re}}
\newcommand{\I}{\mathrm{Im}}

\begin{document}

\preprint{APS/123-QED}

\title{Unique properties of the optical activity in noncentrosymmetric superconductors: \\ sum rule, missing area, and relation with the superconducting Edelstein effect}

\author{Koki Shinada}
 \altaffiliation[shinada.koki.64w@st.kyoto-u.ac.jp ]{}%Lines break automatically or can be forced with \\
\author{Robert Peters}%
 \affiliation{%
 Department of Physics, Kyoto University, Kyoto 606-8502, Japan}%

\date{\today}

\begin{abstract}
We present general properties of the optical activity in noncentrosymmetric materials, including superconductors. We derive a sum rule of the optical activity in general electric states and show that the summation of the spectrum is zero, which is independent of the details of electric states. The optical activity has a $\delta$-function singularity that vanishes in normal phases. However, the singularity emerges in superconducting phases, corresponding to the Meissner effect in the optical conductivity. The spectrum decreases by the superconducting gap and has a missing area compared to the normal phase. This area is exactly equivalent to the coefficient of the $\delta$-function singularity due to the universal sum rule. Furthermore, the coefficient is exactly equivalent to the superconducting Edelstein effect, which has not yet been observed in experiments. Thus, this measurement of the missing area offers an alternative way to observe the superconducting Edelstein effect.
\end{abstract}

\maketitle

\section{Introduction}
Optical responses are one of the key research topics in condensed matter physics because they offer valuable insights into diverse material characteristics, such as momentum-resolved electric spectral functions using angle-resolved photoemission spectroscopy, symmetry breaking and associated domains using the Kerr effect and the second harmonic generation.
The wide range of optical frequencies, spanning from microwaves to X-rays, enables the investigation of phenomena across an extensive spectrum of energy scales.
Recently, terahertz spectroscopy is also attracting attention because important energy scales exist in this regime in condensed matter physics, such as the superconducting gap and collective excitations of magnets \cite{Nicoletti:16}.

Optical responses have also played an essential role in the research of superconductors. It dates back to the observation of the superconducting gap in thin films of $\mathrm{Pb}$ using the far-infrared ray in 1956, which gave the first evidence of the superconducting gap \cite{PhysRev.104.844}. Furthermore, the optical conductivity has contributed to the identification of the gap symmetry of superconductors and the exact measurement of the superfluid density or the magnetic penetration length through the use of a sum rule. This measurement has been mainly done in high-temperature superconductors \cite{PhysRev.109.1398,PhysRevLett.2.331,doi:10.1142/9789814439688_0005,RevModPhys.77.721,Charnukha_2014}. In recent times, the research area of optical responses in superconductors has been more diverse; the third harmonic generation observing the Higgs mode \cite{doi:10.1146/annurev-conmatphys-031119-050813} and the optical conductivity in noncentrosymmetric superconductors \cite{zhao2017global,PhysRevB.100.220501,PhysRevB.105.024308,PhysRevB.107.024513,yang2019lightwave,PhysRevLett.125.097004,PhysRevLett.124.207003} are energetically studied.

In this work, we will extend the study of optical responses to the optical activity in superconductors. The optical activity represents one of the optical responses, and it originates from the spatial dispersion of the optical conductivity, exhibiting the optical rotation, the dichroism, and the birefringence depending on material symmetries \cite{10.1063/1.1729451,PhysRev.171.1065,landau2013electrodynamics,10.1093/acprof:oso/9780198567271.001.0001,Arima_2008,tokura2018nonreciprocal}. It comprises two categories depending on the existence of the time-reversal symmetry ($\mathcal{T}$). One is the natural optical activity with $\mathcal{T}$-symmetry, and the other is the spatially-dispersive magneto-optical effect or the optical magnetoelectric effect without $\mathcal{T}$-symmetry. Spatial inversion symmetry breaking is necessary for a finite optical activity, and it is observed in various systems including chiral molecules as well as the noncentrosymmetric crystals. Despite the ubiquity of optical activity, theoretical studies of optical activity have mainly been carried out in molecular systems \cite{RevModPhys.9.432,barron_2004,https://doi.org/10.1002/chir.10145,autschbach2011time,https://doi.org/10.1002/hlca.202100154}, and research in solids is not developed to the same level.
The band theory of the optical activity in solids is developed in some works \cite{doi:10.1143/JPSJ.39.1013,PhysRevB.48.1384,PhysRevB.81.094525,PhysRevB.82.245118,PhysRevLett.116.077201,PhysRevB.92.235205,PhysRevLett.122.227402,PhysRevB.106.085413}, and it has been applied to various systems including chiral crystals \cite{PhysRevB.79.075438,PhysRevB.97.035158,rerat2021first,PhysRevB.107.224430}, twisted bilayer graphenes \cite{kim2016chiral,SuárezMorell_2017,PhysRevLett.120.046801,PhysRevB.106.245405,https://doi.org/10.1002/adma.202206141,PhysRevB.107.195141}, and a topological antiferromagnet \cite{ahn2022theory}.
Recently, the optical activity is formulated through the multipole theory in solids, revealing the correspondence with molecular systems \cite{10.21468/SciPostPhys.14.5.118}, and the first-principle calculation is carried out based on this formulation \cite{PhysRevB.107.045201}.

While there has been gradual progress in theoretical studies about the optical activity in the normal phase, the research in noncentrosymmetric superconductors remains largely unexplored, except for few work \cite{PhysRevB.81.094525,PhysRevB.88.134514}. Recently, noncentrosymmetric superconductors have attracted increasing attention \cite{bauer2012non}, because they cause novel superconducting states, such as parity-mixing superconductors, topological superconductors, and helical superconductors with finite momentum Cooper pairs. They, furthermore, display unique magnetoelectric responses and nonreciprocal phenomena due to inversion symmetry breaking, including the superconducting Edelstein effect \cite{PhysRevLett.75.2004}, the magnetochiral anisotropy \cite{PhysRevLett.121.026601,doi:10.1126/sciadv.1602390,PhysRevResearch.2.042046}, and the superconducting diode effect \cite{PhysRevLett.128.037001,doi:10.1073/pnas.2119548119,He_2022,ando2020observation}. In addition, the optical activity will give valuable information about these superconductors due to its uniqueness in inversion-symmetry broken systems.

In this paper, we show the general properties of the optical activity in noncentrosymmetric systems, including superconductors. First, we discuss a sum rule of the optical activity valid in all systems and reveal that the summation does not depend on material details and electric states in Sec.~{\ref{sum_OA}}. Second, we formulate the optical activity using Green's functions and discuss a no-go theorem stating the absence of a $\delta$-function singularity, meaning that optical activity does not appear in equilibrium in Sec.~\ref{OA_normal}. Furthermore, we show a typical optical spectrum calculated in a two-dimensional model, including a Rashba spin-orbit coupling, and confirm this sum rule. Third, we discuss the optical activity in noncentrosymmetric superconductors in Sec.~\ref{OA_SC}. In this section, we formulate the optical activity for superconductors and demonstrate that the no-go theorem is broken. For this reason, the singularity appears, and the optical spectrum of the optical activity is reduced compared to the normal phase. This area is called the \textit{missing area} and is exactly equivalent to the coefficient of the $\delta$-function singularity because of the universal sum rule. Furthermore, we reveal that the missing area is exactly equivalent to the superconducting Edelstein effect, where a magnetization is induced by supercurrents, which has not been experimentally observed. The relation established here by the missing area provides an alternative way to observe this effect. Furthermore, we also calculate the optical activity in a two-dimensional noncentrosymmetric superconductor to verify the typical behavior of the missing area. Finally, we conclude this paper in Sec.~\ref{conclusion}.

\section{sum rule of the optical activity} \label{sum_OA}
In this section, we discuss a sum rule of the optical activity. The optical activity is one of the responses to light and, particularly, is related to the optical rotation and the nonreciprocity due to the inversion symmetry breaking. A similar effect is the magneto-optical effect, which is not included in this paper because inversion symmetry breaking is unnecessary.
The optical activity is theoretically described by a spatially dispersive optical conductivity. When applying an electromagnetic wave, electric current, orbital moments and, also, spin moments interact with the light. These responses are described by a general current-current correlation function where the current operator is conjugate to the electromagnetic vector potential and includes spin moments.

First, as a preparation, we discuss an exact symmetry of the current-current correlation function $\Phi_{\mu \nu}(\q,\omega)$, where $\q$ is the wave number and $\omega$ is the frequency. 
The following relationship holds between this correlation function and its complex conjugate as
\begin{eqnarray}
    \Phi^{*}_{\mu \nu}(\q,\omega) = \Phi_{\mu \nu}(-\q,-\omega). 
\end{eqnarray}
This derivation requires only the hermicity of the current operator (see Appendix \ref{app_sum} for a detailed derivation). Next, we expand this correlation function by the wave number $\q$ and discuss the symmetry of the zeroth order $\Phi_{\mu \nu}(\omega) = \Phi_{\mu \nu}(\bm{0},\omega)$ and the first order term $\Phi_{\mu \nu \lambda}(\omega) = \partial_{q_{\lambda}}\Phi_{\mu \nu}(\bm{0},\omega)$. Separating the correlation function into real and imaginary parts, we find that the following symmetry relations for the frequency hold:
\begin{subequations}
\begin{align}
&\left\{ \,
    \begin{aligned}
    & \R \Phi_{\mu \nu}(-\omega) = +\R \Phi_{\mu \nu}(\omega) \\
    & \I \Phi_{\mu \nu}(-\omega) = -\I \Phi_{\mu \nu}(\omega) 
    \end{aligned}
\right. \label{symmetry_0} \\
&\left\{ \,
    \begin{aligned}
    & \R \Phi_{\mu \nu \lambda}(-\omega) = -\R \Phi_{\mu \nu \lambda}(\omega) \\
    & \I \Phi_{\mu \nu \lambda}(-\omega) = + \I \Phi_{\mu \nu \lambda}(\omega).
    \end{aligned}
\right. \label{symmetry_1}
\end{align}
\end{subequations}
The zeroth order term is even for the real part and odd for the imaginary part, and the opposite relation holds for the first order term.

Next, we derive the sum rule. The spatially dispersive optical conductivity is given by
\begin{eqnarray}
    \sigma_{\mu \nu}(\q,\omega) = \frac{\Phi_{\mu \nu}(\q,\omega) - D_{\mu \nu} }{i (\omega + i \delta)}.
\end{eqnarray}
Here, $D_{\mu \nu}$ is the diamagnetic term, which is real. $\delta = +0$ is an adiabatic factor. Using the equation $\lim_{\delta \to +0} 1/(\omega + i\delta) = \mathscr{P}/\omega - i \pi \delta(\omega)$, the optical conductivity is divided into the real part and the imaginary part as
\begin{subequations}
\begin{align}
&\R \sigma_{\mu \nu}(\q,\omega) \nonumber \\
&= \mathscr{P} \frac{\I \Phi_{\mu \nu}(\q,\omega)}{\omega} -\pi \delta(\omega) (\R \Phi_{\mu \nu}(\q,\omega) - D_{\mu \nu}) \label{cond_real} \\  
&\I \sigma_{\mu \nu}(\q,\omega) \nonumber \\
&= -\mathscr{P} \frac{\R \Phi_{\mu \nu}(\q,\omega) -  D_{\mu \nu}}{\omega} -\pi \delta(\omega) \I \Phi_{\mu \nu}(\q,\omega). \label{cond_imag}
\end{align} \label{cond}
\end{subequations}
Here, $\delta(\omega)$ is the $\delta$-function and $\mathscr{P}$ is the principal value.
Similarly, expanding the optical conductivity by $\q$, the real part of the zeroth order term is an even function of $\omega$ and the imaginary part is odd, which can be seen from Eq.~(\ref{symmetry_0}). The opposite relation is satisfied for the first-order term shown in Eq.~(\ref{symmetry_1}). 
The zeroth order term is the usual optical conductivity, and the first order is called the optical activity finite only in noncentrosymmetric systems and the main quantity in this paper.
For the even functions, we can find the following sum rules. Using the Kramers-Kronig relation, the sum rule for the usual optical conductivity (the zeroth order) is established \cite{doi:10.1143/JPSJ.12.570},
\begin{eqnarray}
\int^{\infty}_0 d\omega \R \sigma_{\mu \nu}(\omega) = \frac{\pi}{2} D_{\mu \nu}. \label{sum_cond}
\end{eqnarray}
Next, the sum rule for the optical activity reads (see Appendix \ref{app_sum} for a detailed derivation)
\begin{eqnarray}
    \int^{\infty}_0 d\omega \I \sigma_{\mu \nu \lambda}(\omega) = 0. \label{sum_activ}
\end{eqnarray}
This relation is the first main result of this paper. This equation means that the summation is zero and has universality due to the independence of material details. This property will be important in the following discussion.
This sum rule is partially derived in molecule systems \cite{RevModPhys.9.432,barron_2004}, which are finite systems, and it has been further extended to infinite systems, crystals \cite{doi:10.1143/JPSJ.39.1013,PhysRevB.48.1384,10.21468/SciPostPhys.14.5.118}. However, the derivation is limited to the noninteracting band theory. On the other hand, Eq.~(\ref{sum_activ}) generalizes the sum rule to systems without any assumption and is also valid for, e.g., interacting systems and superconducting states.

\section{Optical activity in noncentrosymmetric crystals} \label{OA_normal}
In this section, we discuss general properties, such as symmetry constraints and a no-go theorem, and typical behaviors of the optical activity in noninteracting crystals.

\subsection{Symmetry classification for the optical activity: \\ natural optical activity and optical magnetoelectric effect}
Response functions are, in general, constrained by time-reversal symmetry, and this constraint is given by the Onsager reciprocal theorem. The optical conductivity satisfies the reciprocal relation \cite{landau2013electrodynamics}
\begin{eqnarray}
    \sigma_{\mu \nu}(\q,\omega,\bm{M}) = \sigma_{\nu \mu}(-\q,\omega,-\bm{M}).
\end{eqnarray}
Here, $\bm{M}$ is a time-reversal symmetry-breaking term such as an external magnetic field or a magnetization. Thus, the symmetric and antisymmetric parts of the optical activity behave differently for the interchange of the indices $\mu \leftrightarrow \nu$ as \cite{PhysRevB.82.245118,10.21468/SciPostPhys.14.5.118}
\begin{subequations}
\begin{align}
    \sigma^{(S)}_{\mu \nu \lambda}(\omega, \bm{M}) &= - \sigma^{(S)}_{\mu \nu \lambda}(\omega, - \bm{M}) \\
    \sigma^{(A)}_{\mu \nu \lambda}(\omega, \bm{M}) &= + \sigma^{(A)}_{\mu \nu \lambda}(\omega, - \bm{M}).
\end{align}
\end{subequations}
These equations show that the symmetric part $\sigma^{(S)}_{\mu \nu \lambda}$ is odd and the antisymmetric part $\sigma^{(A)}_{\mu \nu \lambda}$ is even under the time-reversal operation $\mathcal{T}$. Thus, the symmetric part needs time-reversal breaking ($\bm{M} \neq 0$), but the antisymmetric part does not. The optical activity is further restricted by the spatial inversion symmetry. The optical activity tensors are odd under the spatial-inversion operation $\mathcal{P}$, therefore, the antisymmetric part vanishes if systems have $\mathcal{PT}$ symmetry. Because of the different symmetry constraints of these parts, they are named differently. The antisymmetric part is called the natural optical activity (NOA) and the symmetric part is called the optical magnetoelectric effect.

The NOA is mainly composed of the optical rotation and the circular dichroism and has been studied for a long time. It dates back to the first observation in 1811 by Arago, showing that a quartz displayed an optical rotation. 
The NOA is often used to distinguish chiral molecules because the enantiomers, which are the mirrored states, exhibit the NOA with opposite signs to the original molecules. Furthermore, the NOA is also active in chiral solids, such as Te and Se \cite{PhysRevLett.5.500}, and twisted bilayer graphene, as we have noted in the introduction.
In the aspect of symmetry, the NOA can appear even in $\mathcal{T}$-symmetric systems and, then, purely reflects the crystal symmetry. The antisymmetric optical activity behaves as a rank-2 axial tensor $\alpha_{\xi \lambda} = \varepsilon_{\mu \nu \xi} \sigma^{(A)}_{\mu \nu \lambda}$ ($\varepsilon_{\mu \nu \xi}$ is a totally antisymmetric tensor), and this tensor is active in gyrotropic point groups (GPGs) \cite{PhysRevB.98.165110,doi.org/10.1002/ijch.202200049}.
GPGs are divided into strong and weak GPGs, and weak GPGs are composed of $\mathrm{C_{3v}}, \mathrm{C_{4v}}, \mathrm{C_{6v}}$. These two GPGs generate different types of the NOA. The optical rotation is active in strong GPGs, however, it does not appear in weak GPGs. On the other hand, weak GPGs display the Voigt-Fedorov dichroism or a specific reflection phenomenon \cite{10.1063/1.433207,Graham:96,PhysRevB.92.241116}. This phenomenon was observed, for example, in CdS with $\mathrm{C_{6v}}$ \cite{Ivchenko1978,IVCHENKO1978345}.
Furthermore, in spin-orbit coupled systems, the NOA includes the optical Edelstein effect, where the AC current induces a dynamical magnetization \cite{PhysRevB.92.075113,doi:10.7566/JPSJ.85.033701}.

The symmetric part can be decomposed into a rank-2 axial tensor $\beta_{\mu \xi} = \varepsilon_{\nu \lambda \xi} \sigma^{(S)}_{\mu \nu \lambda}$ and a rank-3 totally symmetric tensor $\gamma_{\mu \nu \lambda} = \sigma^{(S)}_{\mu \nu \lambda} + \sigma^{(S)}_{\nu \lambda \mu} + \sigma^{(S)}_{\lambda \mu \nu }$. $\beta_{\mu \xi}$ corresponds to the optical magnetoelectric response \cite{Arima_2008}, and it induces, e.g., the directional dichroism and the directional birefringence. The response is observed in the typical magnetoelectric material $\mathrm{Cr_2O_3}$ \cite{doi:10.1080/01411599108203448,Krichevtsov_1993}, and now the magnetoelectric optics allows domain imaging of antiferromagnets \cite{kimura2020imaging,PhysRevB.105.094417,PhysRevResearch.4.043063}.
Furthermore, the optical magnetoelectric response is now widely observed \cite{tokura2018nonreciprocal}, and the response caused by magnons in multiferroic magnets is also reported \cite{PhysRevLett.106.057403,takahashi2012magnetoelectric}.
$\gamma_{\mu \nu \lambda}$ is known to be an electric quadrupole response \cite{PhysRevB.82.245118,10.21468/SciPostPhys.14.5.118}, which also induces the directional dichroism \cite{PhysRevLett.122.227402}.

\subsection{Green's function formula of the optical activity for noninteracting systems} \label{green_normal}
We derive the Green's function formula of the optical activity for the noninteracting systems.
The noninteracting Hamiltonian without electromagnetic wave is
\begin{eqnarray}
    H_{0} = \frac{\bm{p}^2}{2m} + V(\bm{x}) + \frac{1}{4m^2}\Bigl( \frac{\partial V(\bm{x})}{\partial \bm{x}} \times \bm{p} \Bigr) \cdot \bm{\sigma}.
\end{eqnarray}
Here, $\bm{p}$ and $\bm{x}$ are the momentum and position operators, respectively, $m$ is the mass of an electron, $V(\bm{x}) = V(\bm{x} + \bm{a})$ is a periodic potential, and $\bm{\sigma}$ is the Pauli matrix representing the spin degrees of freedom.
This Hamiltonian is diagonalized by the Bloch wave function $\ket{\psi_{n\bm{k}}}$ ($\bm{k}$ is the Bloch wave number and $n$ is the band index) as $H_{0} \ket{\psi_{n\bm{k}}} = \epsilon_{n\bm{k}} \ket{\psi_{n\bm{k}}}$. For the following discussion, we define the Bloch Hamiltonian $H_{\bm{k}} = e^{-i \bm{k} \cdot \bm{x}} H_0 e^{i \bm{k} \cdot \bm{x}}$ and the periodic part of the Bloch function $\ket{u_{n\bm{k}}} = e^{-i \bm{k} \cdot \bm{x}} \ket{\psi_{n\bm{k}}}$.
Then, introducing electromagnetic waves by the vector potential $\bm{A}(\bm{x},t)$, the momentum changes as $\bm{p} \to \bm{p} + e\bm{A}(\bm{x},t)$ ($-e < 0$ is the charge of the electron) and the Zeeman term is added. The first-order perturbed Hamiltonian is given by
\begin{subequations}
\begin{align}
    &H_{A} = \frac{e}{2} \Bigl( \bm{v} \cdot \bm{A}(\bm{x},t) + \bm{A}(\bm{x},t) \cdot \bm{v} \Bigr) \\
    &H_{B} = \frac{g_S \mu_B}{2} (\partial_{\bm{x}} \times \bm{A}(\bm{x},t)) \cdot \bm{\sigma}.
\end{align}
\end{subequations}
Here, $\bm{v} = i [H_0 , \bm{x}]$ is the velocity operator, $g_S = 2.002 \cdots$ is the spin $g$-factor, and $\mu_B = e/2m$ is the Bohr magneton. The generalized current operator $\J(\bm{r})$, conjugate to the vector potential $\bm{A}(\bm{r},t)$ ($\bm{r}$ is just the position coordinate, not the operator), is defined as
\begin{eqnarray}
    \J(\bm{r}) &\equiv& - \frac{\delta (H_A + H_B)}{\delta \bm{A}(\bm{r},t)} \nonumber \\
    &=&
    - \frac{e}{2} \{ \bm{v} , \delta(\bm{r} - \bm{x}) \} + \frac{g_S \mu_B}{2} (\bm{\sigma} \times \partial_{\bm{r}}) \delta(\bm{r} - \bm{x}). \nonumber \\ \label{g_current}
\end{eqnarray}
This current operator is the quantity induced by the interaction with electromagnetic waves. Following the dynamical linear response theory given by the Kubo formula, the current-current correlation function using the Green's functions is given by
\begin{align}
&\Phi_{\mu \nu}(\q,\Omega) \nonumber \\
&= 
\int [d^4k] f(\omega) \mathrm{Tr} \Bigl[
G^{RA}(\bm{k}-, \omega) j^{\mu}_{\bm{k},\bm{q}} G^{R}(\bm{k}+,\omega+\Omega) j^{\nu}_{\bm{k},-\bm{q}} \nonumber \\
&+
G^{A}(\bm{k}-, \omega - \Omega) j^{\mu}_{\bm{k},\bm{q}} G^{RA}(\bm{k}+,\omega) j^{\nu}_{\bm{k},-\bm{q}}
\Bigr]. \label{g_correlation_normal}
\end{align}
Here, $G^{R/A}(\bm{k},\omega) = 1/(\omega - H_{\bm{k}} + \mu \pm i\Gamma)$ is the retarded/advanced Green function, and we define $G^{RA} = G^R - G^A$, $\bm{k}\pm = \bm{k} \pm \bm{q}/2$ and $j^{\mu}_{\bm{k},\q} = -e v^{\mu}_{\bm{k}} - \frac{g_S \mu_B}{2} (i\q \times \bm{\sigma})_{\mu}$. $v^{\mu}_{\bm{k}} = \partial H_{\bm{k}}/\partial k_{\mu}$ is the velocity operator of the Bloch Hamiltonian, $\mu$ is a chemical potential, and $f(\omega) = 1/(e^{\beta \omega} + 1)$ is the Fermi distribution function at temperature $1/\beta$. 
The integral symbol is abbreviated as $\int [d^4k] = \int_{-\infty}^{\infty} d\omega/(2 \pi i) \int_{\mathrm{BZ}} d^3k/(2 \pi)^3$.
We phenomenologically introduce the dissipation effect by assuming a finite $\Gamma$. In this calculation, we neglect the diamagnetic term $D_{\mu \nu}$ because it does not contribute to the optical activity due to the independence of $\q$.
The optical activity is attributed to the first-order term of this correlation function by $\q$, thus, two different contributions appear. One is coming from the spin part in the current operator $j^{\mu}_{\bm{k},\bm{q}}$, and the other is the orbital contribution given by the expansion of the Green's function by $\q$. 
In the following, we focus on the spin contribution for simplicity and only consider crystal symmetries and dimensions where the orbital contribution is absent.
The spin term  is given by
\begin{widetext}
\begin{eqnarray}
&&\Phi_{\mu \nu \lambda}(\Omega) 
=
\frac{ieg_S \mu_B}{2} \int [d^4 k]  f(\omega) \mathrm{Tr} \Bigl[
\varepsilon_{\mu \lambda \theta} 
\Bigl\{ G^{RA}(\bm{k}, \omega) \sigma_{\theta} G^{R}(\bm{k},\omega+\Omega) v^{\nu}_{\bm{k}} 
+ G^{A}(\bm{k}, \omega - \Omega) \sigma_{\theta} G^{RA}(\bm{k},\omega) v^{\nu}_{\bm{k}}
\Bigr\} \nonumber \\
&&\hspace{140pt}
- \varepsilon_{\nu \lambda \theta} \Bigl\{
G^{RA}(\bm{k}, \omega) v^{\mu}_{\bm{k}} G^{R}(\bm{k},\omega+\Omega) \sigma_{\theta}
+
G^{A}(\bm{k}, \omega - \Omega) v^{\mu}_{\bm{k}} G^{RA}(\bm{k},\omega) \sigma_{\theta} \Bigr\}
\Bigr]. \label{oa_normal}
\end{eqnarray}
\end{widetext}
Then, we obtain the formula of the optical activity $\sigma_{\mu \nu \lambda}(\Omega) = \Phi_{\mu \nu \lambda}(\Omega)/i(\Omega + i\delta)$.

\subsection{No-go theorem and typical behaviors of the optical activity} \label{property_normal}
\begin{figure*}[t]
\includegraphics[width=0.5\linewidth]{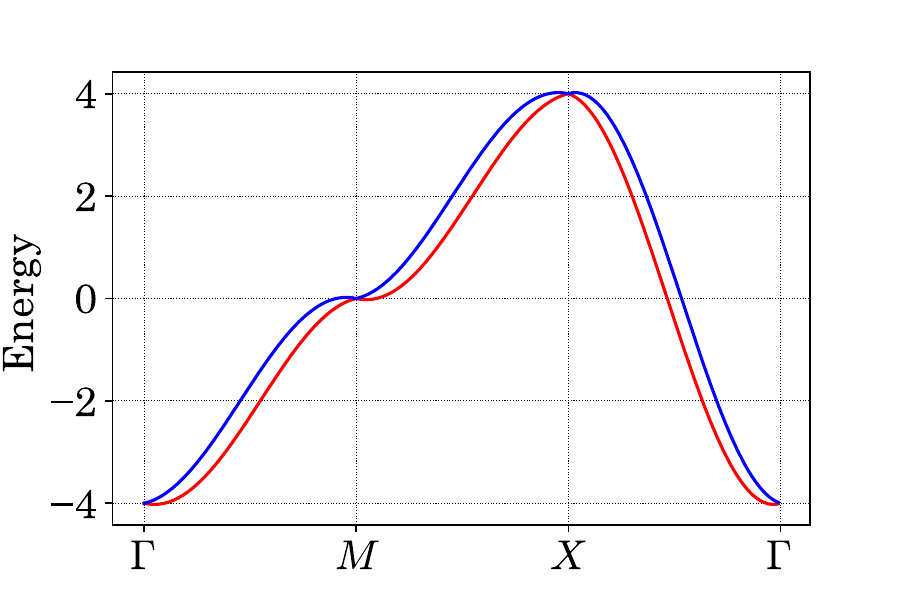}
\includegraphics[width=0.45\linewidth]{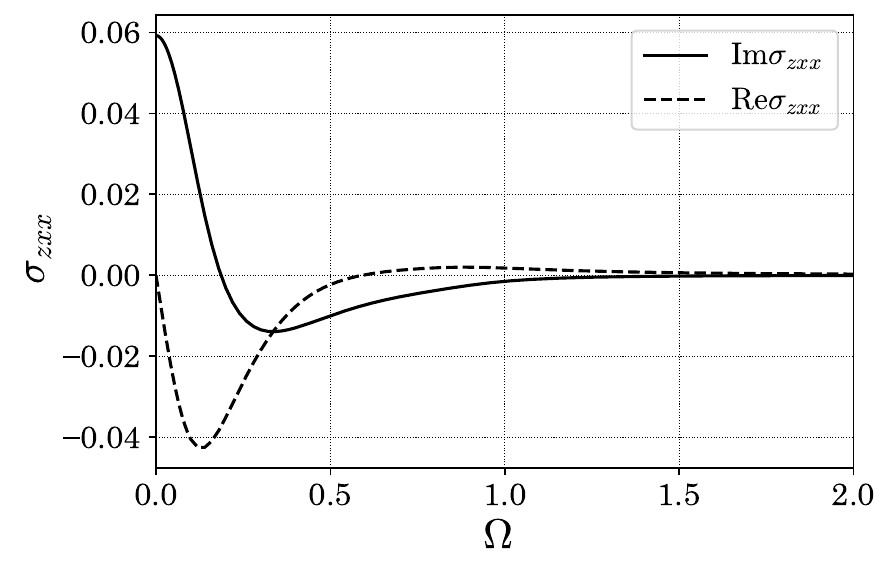}
\caption{(Left) The energy dispersion of the Rashba spin-orbit coupling model (Eq.~\ref{rashba_hamiltonian}).
(Right) Numerical results of the optical activity $\sigma_{zxx}$ for the model Hamiltonian (Eq.~\ref{rashba_hamiltonian}). We set $t=1.0$, $\alpha = 0.3$, $\mu = -0.1$ and $\Gamma = 0.1$. The mesh of the wavenumbers in the BZ is $800 \times 800$. We set Energy and $\Omega$ in units of $t$, and $\sigma_{zxx}$ in units of $e^2 a /\hbar$. In this unit, we assume the electron mass $m \approx 1 \cdot \hbar^2/t a^2$.}  \label{op_normal_fig}
\end{figure*}
We discuss a general property, namely a no-go theorem, using the obtained formula (Eq.~\ref{oa_normal}). The optical activity appears to have a singularity at $\omega = 0$ due to the $\delta$-function in Eqs.~(\ref{cond_real}) and (\ref{cond_imag}). However, we can prove that this singularity vanishes in the normal phase. In fact, taking the limit $\Omega \to 0$, the integrand in Eq.~(\ref{oa_normal}) can be rewritten as a total differential form with respect to the wave number $\bm{k}$ using the fact that $G^{R/A} v^{\lambda}_{\bm{k}} G^{R/A} = \partial_{\lambda} G^{R/A}$, and thus $\Phi_{\mu \nu \lambda}(\Omega = 0) = 0$ resulting in the singularity vanishing. This result shows that the optical activity vanishes in equilibrium. However, this is not the case in superconductors as we will discuss later.

Next, we calculate the optical activity in a simple model and discuss the typical behaviors. We use a 2-dimensional noncentrosymmetric system including the Rashba spin-orbit coupling, however, we note that the optical activity is also expected in other 3-dimensional systems with other types of spin-orbit coupling. This Hamiltonian reads
\begin{eqnarray}
    H_{\bm{k}} = \sum_{\bm{k} \sigma} \epsilon_{\bm{k}} c^{\dagger}_{\sigma \bm{k}} c_{\sigma \bm{k}} + \alpha \sum_{\bm{k} \sigma \sigma'} \bm{g}_{\bm{k}} \cdot \bm{\sigma}_{\sigma \sigma'} c^{\dagger}_{\sigma \bm{k}} c_{\sigma' \bm{k}}. \label{rashba_hamiltonian} 
\end{eqnarray}
Here, we define $\epsilon_{\bm{k}} = -2t( \cos k_x + \cos k_y )$ and $\bm{g}_{\bm{k}} = (\sin k_y , - \sin k_x , 0)$. In this paper, we set the lattice constant $a=1$.
This model is $\mathcal{T}$-symmetric and belongs to the weak gyrotropic point group $\mathrm{C_{4v}}$. Thus, the symmetric part of the optical activity vanishes, however, the antisymmetric part is not forbidden. For this symmetry, there is only one finite component $\sigma_{yzy} = -\sigma_{zyy} = -\sigma_{zxx} = \sigma_{xzx}$. 

Figure~\ref{op_normal_fig} (Left) shows the energy dispersion and there is a band splitting due to the spin-orbit coupling.
Figure~\ref{op_normal_fig} (Right) shows the frequency dependence of the optical activity $\sigma_{zxx}$ at zero temperature. In the numerical calculation, we set $t=1.0$, $\alpha = 0.3$, $\mu = -0.1$, and $\Gamma = 0.1$.
There is a gap of about $0.1 \sim 0.3$ magnitude around the chemical potential $\mu = -0.1$ for direct transitions, not changing the wave numbers $\bm{k}$, as seen in Fig.~\ref{op_normal_fig} (Left). At the frequency $\Omega$ equivalent to the gap, the imaginary part of the optical activity changes its sign. The frequency dependence, including the sign change, can be explained by the concept of intraband and interband transitions. The formula of the imaginary part of the optical activity in the band representation is given by 
\begin{align}
&\I \sigma^{\mathrm{intra}(A)}_{zxx}(\Omega) = \frac{-e g_S \mu_B}{2} \frac{\tau}{(\Omega \tau)^2 + 1} \sum_{n\bm{k}} \frac{\partial f(\tilde{\epsilon}_{n\bm{k}})}{\partial k_x} \sigma^{y}_{nn} \nonumber \\
&\I \sigma^{\mathrm{inter}(A)}_{zxx}(\Omega) \nonumber \\
&= \frac{e g_S \mu_B}{2} \sum_{n\neq m, \bm{k}} \frac{f_{mn\bm{k}}}{\epsilon_{mn\bm{k}}} \frac{\tau}{\tau^2(\epsilon_{mn\bm{k}} - \Omega)^2 + 1} \R \Bigl[ v^x_{\bm{k} mn} \sigma^y_{nm} \Bigr]. \nonumber \\ \label{intra_inter}
\end{align}
This equation is composed of two parts, the intraband effect and the interband effect.
Here, we define the matrix element $M_{mn} = \bra{u_{m\bm{k}}} M \ket{u_{n\bm{k}}}$, $f_{mn\bm{k}} = f(\tilde{\epsilon}_{m\bm{k}}) - f(\tilde{\epsilon}_{n\bm{k}})$, $\epsilon_{mn\bm{k}} = \epsilon_{m\bm{k}} - \epsilon_{n\bm{k}}$, and $\tilde{\epsilon}_{n\bm{k}} = \epsilon_{n\bm{k}} - \mu$.
The intraband effect is attributed to the Fermi surface and behaves like the Drude form $\sim \tau/(1 + (\Omega \tau)^2)$ as seen in Eq.~(\ref{intra_inter}). This behavior can be seen at low frequencies in Fig.~\ref{op_normal_fig}. On the other hand, the interband effect is enhanced at the frequency resonant with the band gap ($\sim 0.2$ in the current model) as seen in Eq.~(\ref{intra_inter}), and Fig.~\ref{op_normal_fig} shows that the sign of the spectrum changes at the corresponding frequency $\Omega \sim 0.2$.
We can confirm that the interband term provides this sign change due to the universal sum rule. 
The low-frequency peak originating from the intraband effect has a constant sign because of the Drude form. On the other hand, the interband effect needs to show a spectrum with an opposite sign so as to cancel the spectrum of the intraband effect and fulfill the universal sum rule (the summation is zero in Eq.~\ref{sum_activ}). Thus, the interband term generates the high-frequency peak with the opposite sign.
Furthermore, in the numerical result in Fig.~\ref{op_normal_fig} (Right), we can confirm the sum rule, and, in fact, the summation of the area is zero ($\sim - 0.0001917\cdots$).

\section{Optical activity in noncentrosymmetric superconductors} \label{OA_SC}
As discussed in previous studies \cite{PhysRevB.81.094525,PhysRevB.88.134514}, a superconducting gyrotropic current changes the frequency dependence of the optical rotation. In this section, we derive the optical activity in superconductors using Green's functions and discuss general properties, including the sum rule and the missing area. In addition, we confirm these properties by a model calculation.

\subsection{Green function formula of the optical activity for superconductors}
We formulate the optical activity for superconductors with Green's functions. In this paper, the superconducting state is treated in the mean-field approximation, and it is described by the BdG Hamiltonian in the band representation
\begin{subequations}
\begin{align}
 &H_{\mathrm{BdG}} = \frac{1}{2} \sum_{\bm{k} nm} \psi^{\dagger}_{n\bm{k}} H^{\mathrm{BdG}}_{\bm{k}nm} \psi_{m \bm{k}}, \\
 &H^{\mathrm{BdG}}_{\bm{k}} =
 \begin{pmatrix}
     H_{\bm{k}} - \mu & - \Delta_{\bm{k}} \\
     - \Delta_{\bm{k}}^{\dagger} & - H_{-\bm{k}}^{\mathrm{T}} + \mu \\
 \end{pmatrix}.
 \end{align}
\end{subequations}
Here, $\bm{\psi}_{\bm{k}}^{\dagger}=(c_{1\bm{k}}^{\dagger}, \cdots, c_{N\bm{k}}^{\dagger}, c_{1 -\bm{k}}, \cdots, c_{N -\bm{k}})$ is the Nambu spinor, $\Delta_{\bm{k}}$ is the pair potential, which is the order parameter of superconductors. We define the transpose of a matrix $M$ as $M^{\mathrm{T}}$ and the hermitian conjugate of a matrix $M$ as $M^{\dagger}$. The current-current correlation function for the generalized current operator (Eq.~\ref{g_current}) is given by
\begin{widetext}
\begin{eqnarray}
\Phi_{\mu \nu}(\q,\Omega) 
&=& 
\frac{1}{2} \int [d^4 k] f(\omega) 
\mathrm{Tr} \Bigl[
G^{RA}_{\mathrm{BdG}}(\bm{k}-, \omega) \tilde{j}^{\mu}_{\bm{k},\bm{q}} G^{R}_{\mathrm{BdG}}(\bm{k}+,\omega+\Omega) \tilde{j}^{\nu}_{\bm{k},-\bm{q}} 
+
G^{A}_{\mathrm{BdG}}(\bm{k}-, \omega - \Omega) \tilde{j}^{\mu}_{\bm{k},\bm{q}} G^{RA}_{\mathrm{BdG}}(\bm{k}+,\omega) \tilde{j}^{\nu}_{\bm{k},-\bm{q}}
\Bigr]. \nonumber \\
\end{eqnarray}
There are some differences from the formula for the normal state (Eq.~\ref{g_correlation_normal}). First, $G^{R/A}_{\mathrm{BdG}}(\bm{k},\omega) = 1/(\omega - H^{\mathrm{BdG}}_{\bm{k}} + \Sigma^{R/A}(\omega) )$ is the Green's function for the BdG Hamiltonian, and $\Sigma^{R/A}(\omega)$ represents the self-energy of the dissipation effect. A specific form will be introduced later. Second, the current operator is expanded to the particle-hole Hilbert space as
\begin{eqnarray}
    \tilde{j}^{\mu}_{\bm{k},\bm{q}}
    =
    \begin{pmatrix}
    j^{\mu}_{\bm{k},\bm{q}} & 0 \\
    0 & -(j^{\mu}_{-\bm{k},\bm{q}})^{\mathrm{T}} \\
    \end{pmatrix}.
\end{eqnarray}
Third, the prefactor $1/2$ is introduced to prevent a double counting of the particle and hole degrees of freedom. After a Taylor expansion by $\q$, the first-order coefficient is decomposed into a spin contribution and an orbital contribution. In this paper, we focus on the spin contribution, which is given by
\begin{eqnarray}
&&\Phi_{\mu \nu \lambda}(\Omega) 
=
\frac{ieg_S \mu_B}{4} \int [d^4 k]  f(\omega) \mathrm{Tr} \Bigl[
\varepsilon_{\mu \lambda \theta} 
\Bigl\{ G^{RA}_{\mathrm{BdG}}(\bm{k}, \omega) \tilde{\sigma}_{\theta} G^{R}_{\mathrm{BdG}}(\bm{k},\omega+\Omega) \tilde{v}^{\nu}_{\bm{k}} 
+ G^{A}_{\mathrm{BdG}}(\bm{k}, \omega - \Omega) \tilde{\sigma}_{\theta} G^{RA}_{\mathrm{BdG}}(\bm{k},\omega) \tilde{v}^{\nu}_{\bm{k}}
\Bigr\} \nonumber \\
&&\hspace{130pt}
- \varepsilon_{\nu \lambda \theta} \Bigl\{
G^{RA}_{\mathrm{BdG}}(\bm{k}, \omega) \tilde{v}^{\mu}_{\bm{k}} G^{R}_{\mathrm{BdG}}(\bm{k},\omega+\Omega) \tilde{\sigma}_{\theta}
+
G^{A}_{\mathrm{BdG}}(\bm{k}, \omega - \Omega) \tilde{v}^{\mu}_{\bm{k}} G^{RA}_{\mathrm{BdG}}(\bm{k},\omega) \tilde{\sigma}_{\theta} \Bigr\}
\Bigr]
\end{eqnarray}
\end{widetext}
Here, the spin operator $\tilde{\sigma}_{\theta}$ and the velocity operator in the Bloch basis $\tilde{v}^{\mu}_{\bm{k}}$ are different from the normal states. They are defined as
\begin{equation}
    \tilde{\sigma}_{\theta}
    =
    \begin{pmatrix}
        \sigma_{\theta} & 0 \\
        0 & - \sigma^{\mathrm{T}}_{\theta} \\
    \end{pmatrix}, \hspace{10pt}
    \tilde{v}^{\mu}_{\bm{k}}
    =
    \begin{pmatrix}
        v^{\mu}_{\bm{k}} & 0 \\
        0 & - (v^{\mu}_{-\bm{k}})^{\mathrm{T}} \\
    \end{pmatrix}. \label{spin_vel_sc}
\end{equation}
Then, we obtain the formula of the optical activity $\sigma_{\mu \nu \lambda}(\Omega) = \Phi_{\mu \nu \lambda}(\Omega)/i(\Omega + i\delta)$.

In this section, we mainly focus on the spin contribution. Of course, the orbital part, in general, contributes to the optical activity, and this part is studied in previous works investigating the superconducting orbital Edelstein effect and the superconducting gyrotropic current resulting in an additional correction to the optical rotation \cite{PhysRevB.81.094525,PhysRevB.88.134514,PhysRevResearch.3.L032012}.

\begin{figure*}[t]
\includegraphics[width=0.45\linewidth]{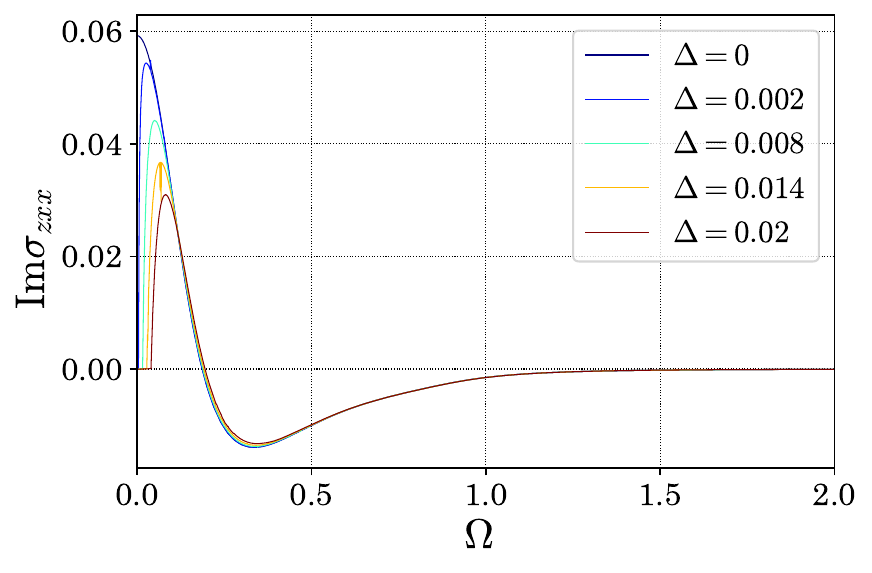}
\includegraphics[width=0.45\linewidth]{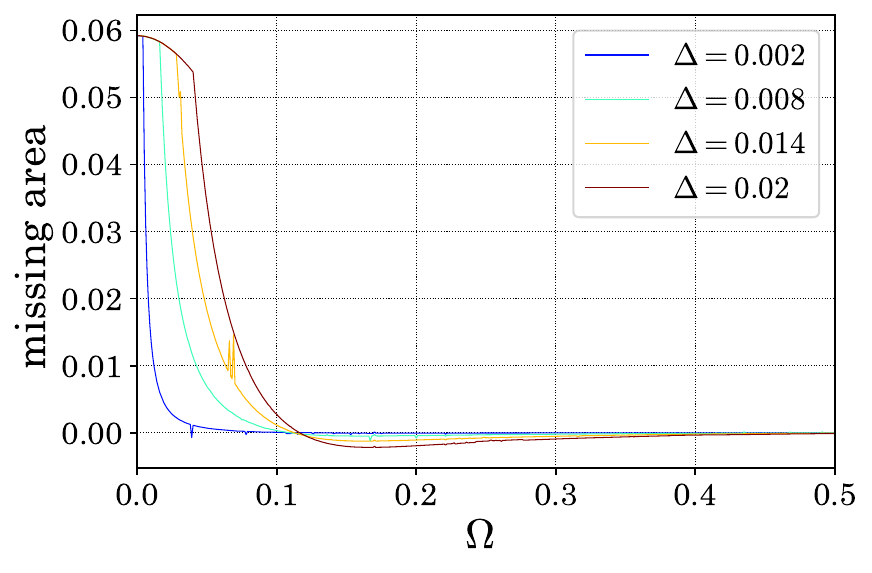}
\caption{(Left) Numerical results of the optical activity in the superconducting phase for several different pair potentials $\Delta$. We set $t=1, \alpha=0.3, \mu=-0.1$, and $\Gamma=0.1$. (Right) Missing area. We plot the difference between the optical activity in the normal phase $\I \sigma_{zxx}^{(n)}$ and in the superconducting phase $\I \sigma_{zxx}^{(s)}$. The integration mesh of the wavenumber in the BZ is $800 \times 800$. We use $\Omega$ and $\Delta$ in units of $t$, and $\sigma_{zxx}$ in units of $e^2 a/\hbar$.} \label{op_sc_fig}
\end{figure*}

\subsection{Relation between the singularity and the superconducting Edelstein effect} \label{edelstein_sc}
As discussed in Sec.~\ref{property_normal}, the singularity due to the $\delta$-function vanishes in the normal state. This result is guaranteed by the identity $G^{R/A} v^{\lambda}_{\bm{k}} G^{R/A} = \partial_{\lambda} G^{R/A}$. However, we will see that a similar identity is not valid in superconducting states. The Bloch velocity operator in superconducting states is described by $\tilde{v}^{\mu}_{\bm{k}}$ (Eq.~\ref{spin_vel_sc}), and this operator does not satisfy $G^{R/A}_{\mathrm{BdG}} \tilde{v}^{\lambda}_{\bm{k}} G^{R/A}_{\mathrm{BdG}} = \partial_{\lambda} G^{R/A}_{\mathrm{BdG}}$, even if the pair potential $\Delta_{\bm{k}}$ is independent of the wavenumber $\bm{k}$. Thus, the integrand in Eq.~(\ref{spin_vel_sc}) cannot be transformed to a total differential form of $\bm{k}$ in the limit of $\Omega \to 0$. This means that the singularity can generally exist in superconducting states. 

A similar singularity appears in the optical conductivity, and the coefficient corresponds to the superfluid density \cite{tinkham2004introduction}. The singularity corresponds to an equilibrium current, which is known to be the Meissner effect. 
Recently, a similar singularity also appears in the nonlinear conductivity resulting in anomalous divergences in the low-frequency regime \cite{PhysRevB.105.024308}, and the origin is the nonreciprocal Meissner effect \cite{PhysRevB.105.L100504}.
The coefficient of the singularity in the optical activity can also be interpreted by a physical effect. The coefficient of the optical activity can be rewritten as
\begin{eqnarray}
    \I \Phi_{\mu \nu \lambda}(0) = \varepsilon_{\mu \lambda \theta} \mathcal{K}_{\nu \theta}
    - (\mu \leftrightarrow \nu).
\end{eqnarray}
$\mathcal{K}_{\nu \theta}$ is the superconducting Edelstein response coefficient, where the supercurrent induces a magnetization in noncentrosymmetric superconductors ($S_{\nu} = \mathcal{K}_{\nu \theta} A_{\theta}$ or $J_{\nu} = \mathcal{K}_{\nu \theta} B_{\theta}$). This response is firstly derived by Edelstein in polar superconductors with Rashba spin-orbit coupling \cite{PhysRevLett.75.2004}, and  subsequent works \cite{PhysRevLett.87.037004,PhysRevB.65.144508,PhysRevB.72.024515,PhysRevB.77.054515} study it in more details. Although the superconducting Edelstein effect is an important response originating in the uniqueness of noncentrosymmetric superconductors, it has not yet been observed in experiments.

\subsection{Missing area} \label{missing_area}
The singularity due to the $\delta$-function is difficult to be directly observed in optical responses. However, it can be exactly measured with the help of the sum rules (Eqs.~\ref{sum_cond} and \ref{sum_activ}). These sum rules state that the summation of the spectrum of the optical responses is independent of material details and does not change before and after superconducting transitions. Thus, the regular part in the superconducting state, accessible in optical measurements, appears to have a reduced area. This is called the missing area and is absorbed in the contribution to the $\delta$-function. This exact relation is used to measure the superfluid density and the penetration length by the optical conductivity. 
This sum rule is called the Ferrell-Glover-Tinkham (FGT) sum rule \cite{PhysRev.109.1398,PhysRevLett.2.331},
and the exact measurement using this sum rule is mainly discussed in high-temperature superconductors \cite{doi:10.1142/9789814439688_0005,RevModPhys.77.721,Charnukha_2014}. 

The discussion of the missing area can be extended to the optical activity. The optical activity also satisfies the sum rule (Eq.~\ref{sum_activ}), which does not change before and after phase transitions. 
Thus, the following equation is established:
\begin{eqnarray}
    \int_{+0}^{\infty} d\Omega \Bigl( \I \sigma^{(n)}_{\mu \nu \lambda}(\Omega) - \I \sigma^{(s)}_{\mu \nu \lambda}(\Omega) \Bigr)
    =
    -\frac{\pi}{2} \I \Phi^{(s)}_{\mu \nu \lambda}(0). \label{sum_rule_missing}
\end{eqnarray}
Here, we label $(n)$ and $(s)$ for the normal state and the superconducting state, respectively.
The left-hand side of this equation represents the missing area, i.e., the difference between the spectral summations of the normal state and the superconducting state at finite frequencies, which is accessible in optical measurements. The right-hand side represents the coefficient of the $\delta$-function singularity. As discussed in the no-go theorems, the singularity is finite only in the superconducting state. Thus, the right-hand side includes only the coefficient of the superconducting state.
As discussed in Sec.~\ref{edelstein_sc}, the coefficient of the $\delta$-function is equivalent to the superconducting Edelstein response that has not yet been observed in experiments. Thus, the missing area measurement gives an alternative way to experimentally determine the superconducting Edelstein effect.

Furthermore, we can directly determine the superconducting Edelstein effect only using the optical spectrum in the superconducting phase. The sum rule states that the summation is zero. Thus, the following equation is established:
\begin{eqnarray}
    \int^{\infty}_{+0} d \Omega \I \sigma^{(s)}_{\mu \nu \lambda}(\Omega) = \frac{\pi}{2} \I \Phi^{(s)}_{\mu \nu \lambda}(0).
\end{eqnarray}

\subsection{Model calculation for the optical activity in a noncentrosymmetric superconductor}
We analyze an optical spectrum of the optical activity in superconductors to verify the above discussion and the missing area. We consider the same model used in Sec.~\ref{property_normal} as the normal Hamiltonian $H_{\bm{k}}$, and the superconducting paring is uniform singlet. Thus, we set $\Delta_{\bm{k}} = i \Delta \sigma_{y}$, where $\Delta$ is real. This model is one of the noncentrosymmetric superconductors including the Rashba spin-orbit coupling. Such superconductors are discussed in the surface atomic-layer superconductors on substrates \cite{uchihashi2021surface} such as the monolayer of FeSe \cite{doi:10.1146/annurev-conmatphys-031016-025242,zakeri2023direct} and noncentrosymmetric bulk superconductors including heavy fermion superconductors with large spin-orbit coupling \cite{bauer2012non}. 

We plot the optical spectrum of the optical activity at various magnitudes of the pair potential $\Delta$ in Fig.~\ref{op_sc_fig}. In this calculation, we phenomenologically introduce the dissipation effect by multiplying a factor $\eta_{\omega}$ obtained by the first Born approximation \cite{10.1093/acprof:oso/9780198507888.001.0001} in the retarded Green's function as
\begin{eqnarray}
    &&G^{R}_{\mathrm{BdG}}(\bm{k},\omega) = \frac{1}{ \eta_{\omega} \omega - H^{(n)}_{\bm{k}} - \eta_{\omega}  \Delta \rho_y \sigma_y} \\
    &&\eta_{\omega} = 1 + \Gamma \biggl(  
    \frac{\theta(|\Delta|-|\omega|)}{\sqrt{\Delta^2 - \omega^2}}
    +
    \frac{i \mathrm{sign}(\omega) \theta(|\omega|-|\Delta|)}{\sqrt{\omega^2 - \Delta^2}}
    \biggr). \nonumber \\
\end{eqnarray}
Here, $H^{(n)}_{\bm{k}}$ is the normal part of $H^{\mathrm{BdG}}_{\bm{k}}$. $\bm{\rho}$ is the Pauli matrix representing the particle-hole Hilbert space, $\theta(x)$ is the step function, and $\mathrm{sign}(x)$ is the sign function returning $+1$ if $x>0$ and $-1$ if $x<0$.
This dissipation effect is consistent with the normal phase introduced in Sec.~\ref{green_normal} in the limit $\Delta \to 0$.
Figure~\ref{op_sc_fig} (Left) shows the optical spectrum, and the weight vanishes for $\omega < 2\Delta$. The curve asymptotically approaches the curve of the normal phase at the high-frequency regime because the superconducting hybridization becomes small.
Figure~\ref{op_sc_fig} (Right) shows the difference between the spectra of the normal state and the superconducting state for visibility. 
Figure~\ref{op_sc_fig} demonstrates that the origin of the missing area comes from the superconducting gap, and the area disappears in the high-frequency regime. Therefore, when measuring this area, it is sufficient to observe it for small frequencies corresponding to the energy scale of the superconducting gap.

\begin{figure}[t]
\includegraphics[width=1\linewidth]{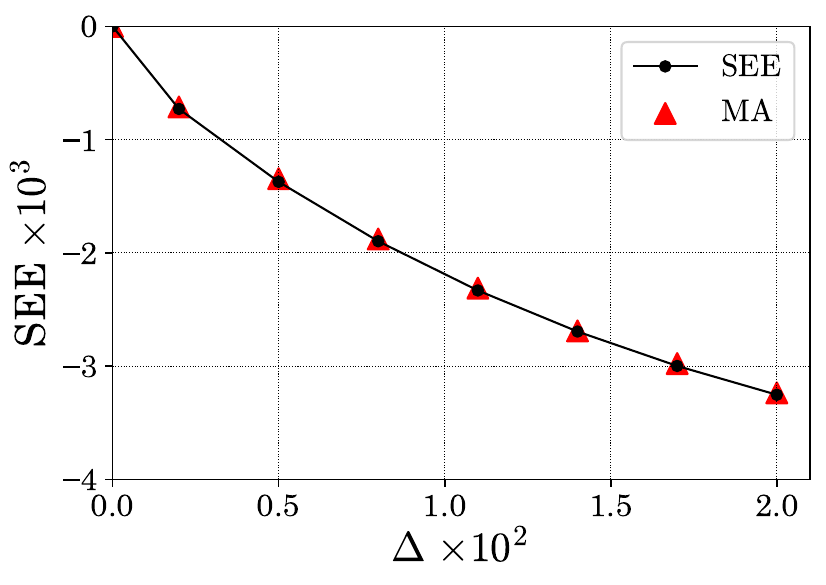}
\caption{Superconducting Edelstein effect (SEE) $\pi \I \Phi_{zxx}(0)/2$ and the missing area (MA) $\int_0^{\Omega_c} d\Omega \I \sigma^{(s)}_{zxx}(\Omega)$. $\Omega_c$ is the cut-off of the integration and we set $\Omega_c = 2.0$. We set $\Delta$ in units of $t$ and SEE in units of $e^2 a /\hbar t$.} \label{missing_fig}
\end{figure}

As discussed in Sec.~\ref{missing_area}, the missing area offers the exact measurement of the superconducting Edelstein effect. In this numerical calculation, we can confirm that the relation is valid as seen in Fig.~\ref{missing_fig}, where we plot both the missing area and the direct calculation of the superconducting Edelstein effect, and we can see that the superconducting Edelstein effect and the missing area coincide.

\section{conclusion} \label{conclusion}
We have investigated general properties of the optical activity in noncentrosymmetric systems, including superconductors. We have derived the sum rule of the optical activity as the property of a two-body correlation function, applicable to general electric states such as interacting systems and superconductors. We have found that the summation is zero, independent of material details in Sec.~\ref{sum_OA}. We have discussed the typical behaviors of the optical activity in the normal phase in Sec.~\ref{OA_normal}. We have formulated the optical activity using Green's functions and discussed a no-go theorem using the obtained formula. The no-go theorem states that the singularity from the $\delta$-function at the zero frequency $\Omega = 0$ is absent, which means that the equilibrium current is forbidden in normal states. In addition, we have calculated the spectrum of the optical activity using a model, including the spin-orbit coupling and seen that there were two peaks with opposite signs in the spectrum. One peak appears around the zero frequency $\Omega = 0$, corresponding to the Drude peak ($\sim \tau/(1 + (\Omega \tau)^2 )$) with finite relaxation time $\tau$. This peak originates at the Fermi surface. Another peak appears at high frequencies and is enhanced around the band gap because its origin is the interband effect. The reason why the two peaks have the opposite sign is that the low-frequency peak from the intraband effect has the constant sign, and the high-frequency peak from the interband effect should show a spectrum with an opposite sign so as to cancel it out due to the sum rule (the summation is zero).

Next, we have discussed the optical activity in noncentrosymmetric superconductors in Sec.~\ref{OA_SC}. We have formulated the optical activity in the superconducting state using Green's functions and discussed some properties using this formula. First, we have discussed a no-go theorem. In superconductors, the theorem no longer holds, and the singularity from the $\delta$-function can appear. Second, we have found a characteristic sum rule similar to the FGT sum rule. Due to the existence of the singularity and the universal sum rule (Eq.~\ref{sum_activ}), the spectrum of the optical activity in the finite frequency regime is reduced, and this missing area is equivalent to the coefficient of the singularity (Eq.~\ref{sum_rule_missing}). Furthermore, we have shown the coefficient is equivalent to the superconducting Edelstein effect, which has not been observed in experiments since the first proposal by Edelstein. Our result has shown that the exact measurement of the missing area gives the alternative way of the observation of the superconducting Edelstein effect. We have also calculated the optical activity using a specific model of a $s$-wave noncentrosymmetric superconductor with a spin-orbit coupling to investigate its typical spectrum. We have found that the missing area originates from the superconducting gap, and that, in the frequency region beyond the superconducting gap, the spectrum asymptotically approaches the normal phase case. Thus, it is sufficient to measure the missing area in the low frequency range about the superconducting gap.

Finally, we comment on the orbital contribution of the optical activity. In this paper, we have focused on the spin contribution. However, the orbital part exists in more general cases as seen in Refs. \cite{PhysRevB.81.094525,PhysRevB.88.134514,PhysRevResearch.3.L032012}. The discussion on the no-go theorem and the missing area involving the orbital part remains as future works.

\section*{acknowledgements}
K.S. thanks Akira Kofuji, Hiroto Tanaka, and Michiya Chazono for variable discussions.
K.S. acknowledges support as a JSPS research fellow and is supported by JSPS KAKENHI, Grant No.22J23393 and No.22KJ2008.
R.P. is supported by JSPS KAKENHI No.23K03300.
The computer calculations in this work have been done using the facilities of the Supercomputer Center, the Institute for Solid State Physics, the University of Tokyo.

\begin{widetext}
\appendix
\section{Derivation of the sum rule} \label{app_sum}
In this appendix, we derive the sum rule of the optical activity.
At first, we show the symmetry relations (Eqs.~\ref{symmetry_0} and \ref{symmetry_1}). The current-current correlation function, in general, can be written in the Lehmann representation as
\begin{eqnarray}
\Phi_{\mu \nu}(\bm{q},\omega)
&&=
\sum_{lm} \frac{e^{-\beta E_l} - e^{-\beta E_m}}{Z ( \omega + i\delta - E_{lm})} \bra{l} J^{\mu}_{\bm{q}} \ket{m} \bra{m} J^{\nu}_{-\bm{q}} \ket{l}. 
\end{eqnarray}
Here, $Z=\mathrm{Tr} [e^{-\beta H}]$ is the partition function.
We use exact eigenstates $\ket{n}$ and eigenvalues $E_n$ of a given system Hamiltonian $H$, and the current operator $\J_{\q} = \int d\bm{r} \J(\bm{r}) e^{-i \q \cdot \bm{r}}$. The current operator is Hermitian $\J^{\dagger}(\bm{r}) = \J(\bm{r})$, thus $\J^{\dagger}_{\q} = \J_{-\q}$. The complex conjugate of this correlation function is
\begin{eqnarray}
    \Phi^{*}_{\mu \nu}(\bm{q},\omega) 
&=&
\sum_{lm} \frac{e^{-\beta E_l} - e^{-\beta E_m}}{Z ( \omega - i\delta - E_{lm})} \bra{m} J^{\mu}_{-\bm{q}} \ket{l} \bra{l} J^{\nu}_{\bm{q}} \ket{m} \nonumber \\
&=&
\sum_{lm} \frac{e^{-\beta E_m} - e^{-\beta E_l}}{Z ( \omega - i\delta + E_{lm})} \bra{l} J^{\mu}_{-\bm{q}} \ket{m} \bra{m} J^{\nu}_{\bm{q}} \ket{l} \nonumber \\
&=& \Phi_{\mu \nu}(-\bm{q},-\omega).
\end{eqnarray}
Then, we can derive Eqs.~(\ref{symmetry_0}) and (\ref{symmetry_1}) by expanding this equation in $\q$.

Next, we move on to the derivation of the sum rule. The imaginary part of the optical activity is
\begin{eqnarray}
    \I \sigma_{\mu \nu \lambda}(\omega) = -\mathscr{P} \frac{\R \Phi_{\mu \nu \lambda}(\omega)}{\omega} - \pi \delta(\omega) \I \Phi_{\mu \nu \lambda}(\omega).
\end{eqnarray}
The imaginary part is an even function of $\omega$, while the real part is odd as we can see in Eq.~(\ref{symmetry_1}). Therefore, integrating the imaginary part along the real-$\omega$ axis is given by
\begin{eqnarray}
    \int^{\infty}_{-\infty} d\omega \I \sigma_{\mu \nu \lambda}(\omega) &=& 2 \int^{\infty}_{0} d\omega \I \sigma_{\mu \nu \lambda}(\omega)  \nonumber \\
    &=& - \mathscr{P} \int^{\infty}_{-\infty} d \omega \frac{\R \Phi_{\mu \nu \lambda}(\omega)}{\omega} - \pi \I \Phi_{\mu \nu \lambda}(0) \nonumber \\
    &=& 0.
\end{eqnarray}
Here, we use the Kramers-Kroning relation, which is valid for retarded functions including $\Phi_{\mu \nu}(\q,\omega)$ due to the analyticity in the upper half-plane of the $\omega$-complex plane, at the third equality. Then, we obtain the sum rule (Eq.~\ref{sum_activ}).

\end{widetext}

\bibliography{ref.bib}

\begin{thebibliography}{86}%
\makeatletter
\providecommand \@ifxundefined [1]{%
 \@ifx{#1\undefined}
}%
\providecommand \@ifnum [1]{%
 \ifnum #1\expandafter \@firstoftwo
 \else \expandafter \@secondoftwo
 \fi
}%
\providecommand \@ifx [1]{%
 \ifx #1\expandafter \@firstoftwo
 \else \expandafter \@secondoftwo
 \fi
}%
\providecommand \natexlab [1]{#1}%
\providecommand \enquote  [1]{``#1''}%
\providecommand \bibnamefont  [1]{#1}%
\providecommand \bibfnamefont [1]{#1}%
\providecommand \citenamefont [1]{#1}%
\providecommand \href@noop [0]{\@secondoftwo}%
\providecommand \href [0]{\begingroup \@sanitize@url \@href}%
\providecommand \@href[1]{\@@startlink{#1}\@@href}%
\providecommand \@@href[1]{\endgroup#1\@@endlink}%
\providecommand \@sanitize@url [0]{\catcode `\\12\catcode `\$12\catcode
  `\&12\catcode `\#12\catcode `\^12\catcode `\_12\catcode `\%12\relax}%
\providecommand \@@startlink[1]{}%
\providecommand \@@endlink[0]{}%
\providecommand \url  [0]{\begingroup\@sanitize@url \@url }%
\providecommand \@url [1]{\endgroup\@href {#1}{\urlprefix }}%
\providecommand \urlprefix  [0]{URL }%
\providecommand \Eprint [0]{\href }%
\providecommand \doibase [0]{https://doi.org/}%
\providecommand \selectlanguage [0]{\@gobble}%
\providecommand \bibinfo  [0]{\@secondoftwo}%
\providecommand \bibfield  [0]{\@secondoftwo}%
\providecommand \translation [1]{[#1]}%
\providecommand \BibitemOpen [0]{}%
\providecommand \bibitemStop [0]{}%
\providecommand \bibitemNoStop [0]{.\EOS\space}%
\providecommand \EOS [0]{\spacefactor3000\relax}%
\providecommand \BibitemShut  [1]{\csname bibitem#1\endcsname}%
\let\auto@bib@innerbib\@empty
%</preamble>
\bibitem [{\citenamefont {Nicoletti}\ and\ \citenamefont
  {Cavalleri}(2016)}]{Nicoletti:16}%
  \BibitemOpen
  \bibfield  {author} {\bibinfo {author} {\bibfnamefont {D.}~\bibnamefont
  {Nicoletti}}\ and\ \bibinfo {author} {\bibfnamefont {A.}~\bibnamefont
  {Cavalleri}},\ }\bibfield  {title} {\bibinfo {title} {Nonlinear light--matter
  interaction at terahertz frequencies},\ }\href
  {https://doi.org/10.1364/AOP.8.000401} {\bibfield  {journal} {\bibinfo
  {journal} {Adv. Opt. Photon.}\ }\textbf {\bibinfo {volume} {8}},\ \bibinfo
  {pages} {401} (\bibinfo {year} {2016})}\BibitemShut {NoStop}%
\bibitem [{\citenamefont {Glover}\ and\ \citenamefont
  {Tinkham}(1956)}]{PhysRev.104.844}%
  \BibitemOpen
  \bibfield  {author} {\bibinfo {author} {\bibfnamefont {R.~E.}\ \bibnamefont
  {Glover}}\ and\ \bibinfo {author} {\bibfnamefont {M.}~\bibnamefont
  {Tinkham}},\ }\bibfield  {title} {\bibinfo {title} {Transmission of
  superconducting films at millimeter-microwave and far infrared frequencies},\
  }\href {https://doi.org/10.1103/PhysRev.104.844} {\bibfield  {journal}
  {\bibinfo  {journal} {Phys. Rev.}\ }\textbf {\bibinfo {volume} {104}},\
  \bibinfo {pages} {844} (\bibinfo {year} {1956})}\BibitemShut {NoStop}%
\bibitem [{\citenamefont {Ferrell}\ and\ \citenamefont
  {Glover}(1958)}]{PhysRev.109.1398}%
  \BibitemOpen
  \bibfield  {author} {\bibinfo {author} {\bibfnamefont {R.~A.}\ \bibnamefont
  {Ferrell}}\ and\ \bibinfo {author} {\bibfnamefont {R.~E.}\ \bibnamefont
  {Glover}},\ }\bibfield  {title} {\bibinfo {title} {Conductivity of
  superconducting films: A sum rule},\ }\href
  {https://doi.org/10.1103/PhysRev.109.1398} {\bibfield  {journal} {\bibinfo
  {journal} {Phys. Rev.}\ }\textbf {\bibinfo {volume} {109}},\ \bibinfo {pages}
  {1398} (\bibinfo {year} {1958})}\BibitemShut {NoStop}%
\bibitem [{\citenamefont {Tinkham}\ and\ \citenamefont
  {Ferrell}(1959)}]{PhysRevLett.2.331}%
  \BibitemOpen
  \bibfield  {author} {\bibinfo {author} {\bibfnamefont {M.}~\bibnamefont
  {Tinkham}}\ and\ \bibinfo {author} {\bibfnamefont {R.~A.}\ \bibnamefont
  {Ferrell}},\ }\bibfield  {title} {\bibinfo {title} {Determination of the
  superconducting skin depth from the energy gap and sum rule},\ }\href
  {https://doi.org/10.1103/PhysRevLett.2.331} {\bibfield  {journal} {\bibinfo
  {journal} {Phys. Rev. Lett.}\ }\textbf {\bibinfo {volume} {2}},\ \bibinfo
  {pages} {331} (\bibinfo {year} {1959})}\BibitemShut {NoStop}%
\bibitem [{\citenamefont {Tanner}\ and\ \citenamefont
  {Timusk}()}]{doi:10.1142/9789814439688_0005}%
  \BibitemOpen
  \bibfield  {author} {\bibinfo {author} {\bibfnamefont {D.~B.}\ \bibnamefont
  {Tanner}}\ and\ \bibinfo {author} {\bibfnamefont {T.}~\bibnamefont
  {Timusk}},\ }\bibinfo {title} {Optical properties of high-temperature
  superconductors},\ in\ \href {https://doi.org/10.1142/9789814439688_0005}
  {\emph {\bibinfo {booktitle} {Physical Properties of High Temperature
  Superconductors III}}},\ pp.\ \bibinfo {pages} {363--469}\BibitemShut
  {NoStop}%
\bibitem [{\citenamefont {Basov}\ and\ \citenamefont
  {Timusk}(2005)}]{RevModPhys.77.721}%
  \BibitemOpen
  \bibfield  {author} {\bibinfo {author} {\bibfnamefont {D.~N.}\ \bibnamefont
  {Basov}}\ and\ \bibinfo {author} {\bibfnamefont {T.}~\bibnamefont {Timusk}},\
  }\bibfield  {title} {\bibinfo {title} {Electrodynamics of high-${T}_{c}$
  superconductors},\ }\href {https://doi.org/10.1103/RevModPhys.77.721}
  {\bibfield  {journal} {\bibinfo  {journal} {Rev. Mod. Phys.}\ }\textbf
  {\bibinfo {volume} {77}},\ \bibinfo {pages} {721} (\bibinfo {year}
  {2005})}\BibitemShut {NoStop}%
\bibitem [{\citenamefont {Charnukha}(2014)}]{Charnukha_2014}%
  \BibitemOpen
  \bibfield  {author} {\bibinfo {author} {\bibfnamefont {A.}~\bibnamefont
  {Charnukha}},\ }\bibfield  {title} {\bibinfo {title} {Optical conductivity of
  iron-based superconductors},\ }\href
  {https://doi.org/10.1088/0953-8984/26/25/253203} {\bibfield  {journal}
  {\bibinfo  {journal} {Journal of Physics: Condensed Matter}\ }\textbf
  {\bibinfo {volume} {26}},\ \bibinfo {pages} {253203} (\bibinfo {year}
  {2014})}\BibitemShut {NoStop}%
\bibitem [{\citenamefont {Shimano}\ and\ \citenamefont
  {Tsuji}(2020)}]{doi:10.1146/annurev-conmatphys-031119-050813}%
  \BibitemOpen
  \bibfield  {author} {\bibinfo {author} {\bibfnamefont {R.}~\bibnamefont
  {Shimano}}\ and\ \bibinfo {author} {\bibfnamefont {N.}~\bibnamefont
  {Tsuji}},\ }\bibfield  {title} {\bibinfo {title} {Higgs mode in
  superconductors},\ }\href
  {https://doi.org/10.1146/annurev-conmatphys-031119-050813} {\bibfield
  {journal} {\bibinfo  {journal} {Annual Review of Condensed Matter Physics}\
  }\textbf {\bibinfo {volume} {11}},\ \bibinfo {pages} {103} (\bibinfo {year}
  {2020})}\BibitemShut {NoStop}%
\bibitem [{\citenamefont {Zhao}\ \emph {et~al.}(2017)\citenamefont {Zhao},
  \citenamefont {Belvin}, \citenamefont {Liang}, \citenamefont {Bonn},
  \citenamefont {Hardy}, \citenamefont {Armitage},\ and\ \citenamefont
  {Hsieh}}]{zhao2017global}%
  \BibitemOpen
  \bibfield  {author} {\bibinfo {author} {\bibfnamefont {L.}~\bibnamefont
  {Zhao}}, \bibinfo {author} {\bibfnamefont {C.}~\bibnamefont {Belvin}},
  \bibinfo {author} {\bibfnamefont {R.}~\bibnamefont {Liang}}, \bibinfo
  {author} {\bibfnamefont {D.}~\bibnamefont {Bonn}}, \bibinfo {author}
  {\bibfnamefont {W.}~\bibnamefont {Hardy}}, \bibinfo {author} {\bibfnamefont
  {N.}~\bibnamefont {Armitage}},\ and\ \bibinfo {author} {\bibfnamefont
  {D.}~\bibnamefont {Hsieh}},\ }\bibfield  {title} {\bibinfo {title} {A global
  inversion-symmetry-broken phase inside the pseudogap region of yba2cu3o y},\
  }\href {https://www.nature.com/articles/nphys3962} {\bibfield  {journal}
  {\bibinfo  {journal} {Nature Physics}\ }\textbf {\bibinfo {volume} {13}},\
  \bibinfo {pages} {250} (\bibinfo {year} {2017})}\BibitemShut {NoStop}%
\bibitem [{\citenamefont {Xu}\ \emph {et~al.}(2019)\citenamefont {Xu},
  \citenamefont {Morimoto},\ and\ \citenamefont {Moore}}]{PhysRevB.100.220501}%
  \BibitemOpen
  \bibfield  {author} {\bibinfo {author} {\bibfnamefont {T.}~\bibnamefont
  {Xu}}, \bibinfo {author} {\bibfnamefont {T.}~\bibnamefont {Morimoto}},\ and\
  \bibinfo {author} {\bibfnamefont {J.~E.}\ \bibnamefont {Moore}},\ }\bibfield
  {title} {\bibinfo {title} {Nonlinear optical effects in
  inversion-symmetry-breaking superconductors},\ }\href
  {https://doi.org/10.1103/PhysRevB.100.220501} {\bibfield  {journal} {\bibinfo
   {journal} {Phys. Rev. B}\ }\textbf {\bibinfo {volume} {100}},\ \bibinfo
  {pages} {220501} (\bibinfo {year} {2019})}\BibitemShut {NoStop}%
\bibitem [{\citenamefont {Watanabe}\ \emph
  {et~al.}(2022{\natexlab{a}})\citenamefont {Watanabe}, \citenamefont {Daido},\
  and\ \citenamefont {Yanase}}]{PhysRevB.105.024308}%
  \BibitemOpen
  \bibfield  {author} {\bibinfo {author} {\bibfnamefont {H.}~\bibnamefont
  {Watanabe}}, \bibinfo {author} {\bibfnamefont {A.}~\bibnamefont {Daido}},\
  and\ \bibinfo {author} {\bibfnamefont {Y.}~\bibnamefont {Yanase}},\
  }\bibfield  {title} {\bibinfo {title} {Nonreciprocal optical response in
  parity-breaking superconductors},\ }\href
  {https://doi.org/10.1103/PhysRevB.105.024308} {\bibfield  {journal} {\bibinfo
   {journal} {Phys. Rev. B}\ }\textbf {\bibinfo {volume} {105}},\ \bibinfo
  {pages} {024308} (\bibinfo {year} {2022}{\natexlab{a}})}\BibitemShut
  {NoStop}%
\bibitem [{\citenamefont {Tanaka}\ \emph {et~al.}(2023)\citenamefont {Tanaka},
  \citenamefont {Watanabe},\ and\ \citenamefont
  {Yanase}}]{PhysRevB.107.024513}%
  \BibitemOpen
  \bibfield  {author} {\bibinfo {author} {\bibfnamefont {H.}~\bibnamefont
  {Tanaka}}, \bibinfo {author} {\bibfnamefont {H.}~\bibnamefont {Watanabe}},\
  and\ \bibinfo {author} {\bibfnamefont {Y.}~\bibnamefont {Yanase}},\
  }\bibfield  {title} {\bibinfo {title} {Nonlinear optical responses in
  noncentrosymmetric superconductors},\ }\href
  {https://doi.org/10.1103/PhysRevB.107.024513} {\bibfield  {journal} {\bibinfo
   {journal} {Phys. Rev. B}\ }\textbf {\bibinfo {volume} {107}},\ \bibinfo
  {pages} {024513} (\bibinfo {year} {2023})}\BibitemShut {NoStop}%
\bibitem [{\citenamefont {Yang}\ \emph {et~al.}(2019)\citenamefont {Yang},
  \citenamefont {Vaswani}, \citenamefont {Sundahl}, \citenamefont {Mootz},
  \citenamefont {Luo}, \citenamefont {Kang}, \citenamefont {Perakis},
  \citenamefont {Eom},\ and\ \citenamefont {Wang}}]{yang2019lightwave}%
  \BibitemOpen
  \bibfield  {author} {\bibinfo {author} {\bibfnamefont {X.}~\bibnamefont
  {Yang}}, \bibinfo {author} {\bibfnamefont {C.}~\bibnamefont {Vaswani}},
  \bibinfo {author} {\bibfnamefont {C.}~\bibnamefont {Sundahl}}, \bibinfo
  {author} {\bibfnamefont {M.}~\bibnamefont {Mootz}}, \bibinfo {author}
  {\bibfnamefont {L.}~\bibnamefont {Luo}}, \bibinfo {author} {\bibfnamefont
  {J.}~\bibnamefont {Kang}}, \bibinfo {author} {\bibfnamefont {I.}~\bibnamefont
  {Perakis}}, \bibinfo {author} {\bibfnamefont {C.}~\bibnamefont {Eom}},\ and\
  \bibinfo {author} {\bibfnamefont {J.}~\bibnamefont {Wang}},\ }\bibfield
  {title} {\bibinfo {title} {Lightwave-driven gapless superconductivity and
  forbidden quantum beats by terahertz symmetry breaking},\ }\href
  {https://www.nature.com/articles/s41566-019-0470-y} {\bibfield  {journal}
  {\bibinfo  {journal} {Nature Photonics}\ }\textbf {\bibinfo {volume} {13}},\
  \bibinfo {pages} {707} (\bibinfo {year} {2019})}\BibitemShut {NoStop}%
\bibitem [{\citenamefont {Nakamura}\ \emph {et~al.}(2020)\citenamefont
  {Nakamura}, \citenamefont {Katsumi}, \citenamefont {Terai},\ and\
  \citenamefont {Shimano}}]{PhysRevLett.125.097004}%
  \BibitemOpen
  \bibfield  {author} {\bibinfo {author} {\bibfnamefont {S.}~\bibnamefont
  {Nakamura}}, \bibinfo {author} {\bibfnamefont {K.}~\bibnamefont {Katsumi}},
  \bibinfo {author} {\bibfnamefont {H.}~\bibnamefont {Terai}},\ and\ \bibinfo
  {author} {\bibfnamefont {R.}~\bibnamefont {Shimano}},\ }\bibfield  {title}
  {\bibinfo {title} {Nonreciprocal terahertz second-harmonic generation in
  superconducting nbn under supercurrent injection},\ }\href
  {https://doi.org/10.1103/PhysRevLett.125.097004} {\bibfield  {journal}
  {\bibinfo  {journal} {Phys. Rev. Lett.}\ }\textbf {\bibinfo {volume} {125}},\
  \bibinfo {pages} {097004} (\bibinfo {year} {2020})}\BibitemShut {NoStop}%
\bibitem [{\citenamefont {Vaswani}\ \emph {et~al.}(2020)\citenamefont
  {Vaswani}, \citenamefont {Mootz}, \citenamefont {Sundahl}, \citenamefont
  {Mudiyanselage}, \citenamefont {Kang}, \citenamefont {Yang}, \citenamefont
  {Cheng}, \citenamefont {Huang}, \citenamefont {Kim}, \citenamefont {Liu},
  \citenamefont {Luo}, \citenamefont {Perakis}, \citenamefont {Eom},\ and\
  \citenamefont {Wang}}]{PhysRevLett.124.207003}%
  \BibitemOpen
  \bibfield  {author} {\bibinfo {author} {\bibfnamefont {C.}~\bibnamefont
  {Vaswani}}, \bibinfo {author} {\bibfnamefont {M.}~\bibnamefont {Mootz}},
  \bibinfo {author} {\bibfnamefont {C.}~\bibnamefont {Sundahl}}, \bibinfo
  {author} {\bibfnamefont {D.~H.}\ \bibnamefont {Mudiyanselage}}, \bibinfo
  {author} {\bibfnamefont {J.~H.}\ \bibnamefont {Kang}}, \bibinfo {author}
  {\bibfnamefont {X.}~\bibnamefont {Yang}}, \bibinfo {author} {\bibfnamefont
  {D.}~\bibnamefont {Cheng}}, \bibinfo {author} {\bibfnamefont
  {C.}~\bibnamefont {Huang}}, \bibinfo {author} {\bibfnamefont {R.~H.~J.}\
  \bibnamefont {Kim}}, \bibinfo {author} {\bibfnamefont {Z.}~\bibnamefont
  {Liu}}, \bibinfo {author} {\bibfnamefont {L.}~\bibnamefont {Luo}}, \bibinfo
  {author} {\bibfnamefont {I.~E.}\ \bibnamefont {Perakis}}, \bibinfo {author}
  {\bibfnamefont {C.~B.}\ \bibnamefont {Eom}},\ and\ \bibinfo {author}
  {\bibfnamefont {J.}~\bibnamefont {Wang}},\ }\bibfield  {title} {\bibinfo
  {title} {Terahertz second-harmonic generation from lightwave acceleration of
  symmetry-breaking nonlinear supercurrents},\ }\href
  {https://doi.org/10.1103/PhysRevLett.124.207003} {\bibfield  {journal}
  {\bibinfo  {journal} {Phys. Rev. Lett.}\ }\textbf {\bibinfo {volume} {124}},\
  \bibinfo {pages} {207003} (\bibinfo {year} {2020})}\BibitemShut {NoStop}%
\bibitem [{\citenamefont {Brown}\ \emph {et~al.}(2004)\citenamefont {Brown},
  \citenamefont {Shtrikman},\ and\ \citenamefont {Treves}}]{10.1063/1.1729451}%
  \BibitemOpen
  \bibfield  {author} {\bibinfo {author} {\bibfnamefont {J.}~\bibnamefont
  {Brown}, \bibfnamefont {W.~F.}}, \bibinfo {author} {\bibfnamefont
  {S.}~\bibnamefont {Shtrikman}},\ and\ \bibinfo {author} {\bibfnamefont
  {D.}~\bibnamefont {Treves}},\ }\bibfield  {title} {\bibinfo {title}
  {{Possibility of Visual Observation of Antiferromagnetic Domains}},\ }\href
  {https://doi.org/10.1063/1.1729451} {\bibfield  {journal} {\bibinfo
  {journal} {Journal of Applied Physics}\ }\textbf {\bibinfo {volume} {34}},\
  \bibinfo {pages} {1233} (\bibinfo {year} {2004})}\BibitemShut {NoStop}%
\bibitem [{\citenamefont {Hornreich}\ and\ \citenamefont
  {Shtrikman}(1968)}]{PhysRev.171.1065}%
  \BibitemOpen
  \bibfield  {author} {\bibinfo {author} {\bibfnamefont {R.~M.}\ \bibnamefont
  {Hornreich}}\ and\ \bibinfo {author} {\bibfnamefont {S.}~\bibnamefont
  {Shtrikman}},\ }\bibfield  {title} {\bibinfo {title} {Theory of gyrotropic
  birefringence},\ }\href {https://doi.org/10.1103/PhysRev.171.1065} {\bibfield
   {journal} {\bibinfo  {journal} {Phys. Rev.}\ }\textbf {\bibinfo {volume}
  {171}},\ \bibinfo {pages} {1065} (\bibinfo {year} {1968})}\BibitemShut
  {NoStop}%
\bibitem [{\citenamefont {Landau}\ \emph {et~al.}(2013)\citenamefont {Landau},
  \citenamefont {Bell}, \citenamefont {Kearsley}, \citenamefont {Pitaevskii},
  \citenamefont {Lifshitz},\ and\ \citenamefont
  {Sykes}}]{landau2013electrodynamics}%
  \BibitemOpen
  \bibfield  {author} {\bibinfo {author} {\bibfnamefont {L.~D.}\ \bibnamefont
  {Landau}}, \bibinfo {author} {\bibfnamefont {J.~S.}\ \bibnamefont {Bell}},
  \bibinfo {author} {\bibfnamefont {M.}~\bibnamefont {Kearsley}}, \bibinfo
  {author} {\bibfnamefont {L.}~\bibnamefont {Pitaevskii}}, \bibinfo {author}
  {\bibfnamefont {E.}~\bibnamefont {Lifshitz}},\ and\ \bibinfo {author}
  {\bibfnamefont {J.}~\bibnamefont {Sykes}},\ }\href@noop {} {\emph {\bibinfo
  {title} {Electrodynamics of continuous media}}},\ Vol.~\bibinfo {volume} {8}\
  (\bibinfo  {publisher} {elsevier},\ \bibinfo {year} {2013})\BibitemShut
  {NoStop}%
\bibitem [{\citenamefont {Raab}\ and\ \citenamefont
  {de~Lange}(2004)}]{10.1093/acprof:oso/9780198567271.001.0001}%
  \BibitemOpen
  \bibfield  {author} {\bibinfo {author} {\bibfnamefont {R.~E.}\ \bibnamefont
  {Raab}}\ and\ \bibinfo {author} {\bibfnamefont {O.~L.}\ \bibnamefont
  {de~Lange}},\ }\href
  {https://doi.org/10.1093/acprof:oso/9780198567271.001.0001} {\emph {\bibinfo
  {title} {{Multipole Theory in Electromagnetism: Classical, quantum, and
  symmetry aspects, with applications}}}}\ (\bibinfo  {publisher} {Oxford
  University Press},\ \bibinfo {year} {2004})\BibitemShut {NoStop}%
\bibitem [{\citenamefont {Arima}(2008)}]{Arima_2008}%
  \BibitemOpen
  \bibfield  {author} {\bibinfo {author} {\bibfnamefont {T.}~\bibnamefont
  {Arima}},\ }\bibfield  {title} {\bibinfo {title} {Magneto-electric optics in
  non-centrosymmetric ferromagnets},\ }\href
  {https://doi.org/10.1088/0953-8984/20/43/434211} {\bibfield  {journal}
  {\bibinfo  {journal} {Journal of Physics: Condensed Matter}\ }\textbf
  {\bibinfo {volume} {20}},\ \bibinfo {pages} {434211} (\bibinfo {year}
  {2008})}\BibitemShut {NoStop}%
\bibitem [{\citenamefont {Tokura}\ and\ \citenamefont
  {Nagaosa}(2018)}]{tokura2018nonreciprocal}%
  \BibitemOpen
  \bibfield  {author} {\bibinfo {author} {\bibfnamefont {Y.}~\bibnamefont
  {Tokura}}\ and\ \bibinfo {author} {\bibfnamefont {N.}~\bibnamefont
  {Nagaosa}},\ }\bibfield  {title} {\bibinfo {title} {Nonreciprocal responses
  from non-centrosymmetric quantum materials},\ }\href
  {https://www.nature.com/articles/s41467-018-05759-4#citeas} {\bibfield
  {journal} {\bibinfo  {journal} {Nature communications}\ }\textbf {\bibinfo
  {volume} {9}},\ \bibinfo {pages} {3740} (\bibinfo {year} {2018})}\BibitemShut
  {NoStop}%
\bibitem [{\citenamefont {Condon}(1937)}]{RevModPhys.9.432}%
  \BibitemOpen
  \bibfield  {author} {\bibinfo {author} {\bibfnamefont {E.~U.}\ \bibnamefont
  {Condon}},\ }\bibfield  {title} {\bibinfo {title} {Theories of optical
  rotatory power},\ }\href {https://doi.org/10.1103/RevModPhys.9.432}
  {\bibfield  {journal} {\bibinfo  {journal} {Rev. Mod. Phys.}\ }\textbf
  {\bibinfo {volume} {9}},\ \bibinfo {pages} {432} (\bibinfo {year}
  {1937})}\BibitemShut {NoStop}%
\bibitem [{\citenamefont {Barron}(2004)}]{barron_2004}%
  \BibitemOpen
  \bibfield  {author} {\bibinfo {author} {\bibfnamefont {L.~D.}\ \bibnamefont
  {Barron}},\ }\href {https://doi.org/10.1017/CBO9780511535468} {\emph
  {\bibinfo {title} {Molecular Light Scattering and Optical Activity}}},\
  \bibinfo {edition} {2nd}\ ed.\ (\bibinfo  {publisher} {Cambridge University
  Press},\ \bibinfo {year} {2004})\BibitemShut {NoStop}%
\bibitem [{\citenamefont
  {Polavarapu}(2002)}]{https://doi.org/10.1002/chir.10145}%
  \BibitemOpen
  \bibfield  {author} {\bibinfo {author} {\bibfnamefont {P.~L.}\ \bibnamefont
  {Polavarapu}},\ }\bibfield  {title} {\bibinfo {title} {Optical rotation:
  Recent advances in determining the absolute configuration},\ }\href
  {https://doi.org/https://doi.org/10.1002/chir.10145} {\bibfield  {journal}
  {\bibinfo  {journal} {Chirality}\ }\textbf {\bibinfo {volume} {14}},\
  \bibinfo {pages} {768} (\bibinfo {year} {2002})}\BibitemShut {NoStop}%
\bibitem [{\citenamefont {Autschbach}\ \emph {et~al.}(2011)\citenamefont
  {Autschbach}, \citenamefont {Nitsch-Velasquez},\ and\ \citenamefont
  {Rudolph}}]{autschbach2011time}%
  \BibitemOpen
  \bibfield  {author} {\bibinfo {author} {\bibfnamefont {J.}~\bibnamefont
  {Autschbach}}, \bibinfo {author} {\bibfnamefont {L.}~\bibnamefont
  {Nitsch-Velasquez}},\ and\ \bibinfo {author} {\bibfnamefont {M.}~\bibnamefont
  {Rudolph}},\ }\bibfield  {title} {\bibinfo {title} {Time-dependent density
  functional response theory for electronic chiroptical properties of chiral
  molecules},\ }\href {https://link.springer.com/chapter/10.1007/128_2010_72}
  {\bibfield  {journal} {\bibinfo  {journal} {Electronic and Magnetic
  Properties of Chiral Molecules and Supramolecular Architectures}\ ,\ \bibinfo
  {pages} {1}} (\bibinfo {year} {2011})}\BibitemShut {NoStop}%
\bibitem [{\citenamefont {Mattiat}\ and\ \citenamefont
  {Luber}(2021)}]{https://doi.org/10.1002/hlca.202100154}%
  \BibitemOpen
  \bibfield  {author} {\bibinfo {author} {\bibfnamefont {J.}~\bibnamefont
  {Mattiat}}\ and\ \bibinfo {author} {\bibfnamefont {S.}~\bibnamefont
  {Luber}},\ }\bibfield  {title} {\bibinfo {title} {Recent progress in the
  simulation of chiral systems with real time propagation methods},\ }\href
  {https://doi.org/https://doi.org/10.1002/hlca.202100154} {\bibfield
  {journal} {\bibinfo  {journal} {Helvetica Chimica Acta}\ }\textbf {\bibinfo
  {volume} {104}},\ \bibinfo {pages} {e2100154} (\bibinfo {year}
  {2021})}\BibitemShut {NoStop}%
\bibitem [{\citenamefont {Natori}(1975)}]{doi:10.1143/JPSJ.39.1013}%
  \BibitemOpen
  \bibfield  {author} {\bibinfo {author} {\bibfnamefont {K.}~\bibnamefont
  {Natori}},\ }\bibfield  {title} {\bibinfo {title} {Band theory of the optical
  activity of crystals},\ }\href {https://doi.org/10.1143/JPSJ.39.1013}
  {\bibfield  {journal} {\bibinfo  {journal} {Journal of the Physical Society
  of Japan}\ }\textbf {\bibinfo {volume} {39}},\ \bibinfo {pages} {1013}
  (\bibinfo {year} {1975})}\BibitemShut {NoStop}%
\bibitem [{\citenamefont {Zhong}\ \emph {et~al.}(1993)\citenamefont {Zhong},
  \citenamefont {Levine}, \citenamefont {Allan},\ and\ \citenamefont
  {Wilkins}}]{PhysRevB.48.1384}%
  \BibitemOpen
  \bibfield  {author} {\bibinfo {author} {\bibfnamefont {H.}~\bibnamefont
  {Zhong}}, \bibinfo {author} {\bibfnamefont {Z.~H.}\ \bibnamefont {Levine}},
  \bibinfo {author} {\bibfnamefont {D.~C.}\ \bibnamefont {Allan}},\ and\
  \bibinfo {author} {\bibfnamefont {J.~W.}\ \bibnamefont {Wilkins}},\
  }\bibfield  {title} {\bibinfo {title} {Band-theoretic calculations of the
  optical-activity tensor of \ensuremath{\alpha}-quartz and trigonal
  $\mathrm{Se}$},\ }\href {https://doi.org/10.1103/PhysRevB.48.1384} {\bibfield
   {journal} {\bibinfo  {journal} {Phys. Rev. B}\ }\textbf {\bibinfo {volume}
  {48}},\ \bibinfo {pages} {1384} (\bibinfo {year} {1993})}\BibitemShut
  {NoStop}%
\bibitem [{\citenamefont {Mineev}\ and\ \citenamefont
  {Yoshioka}(2010)}]{PhysRevB.81.094525}%
  \BibitemOpen
  \bibfield  {author} {\bibinfo {author} {\bibfnamefont {V.~P.}\ \bibnamefont
  {Mineev}}\ and\ \bibinfo {author} {\bibfnamefont {Y.}~\bibnamefont
  {Yoshioka}},\ }\bibfield  {title} {\bibinfo {title} {Optical activity of
  noncentrosymmetric metals},\ }\href
  {https://doi.org/10.1103/PhysRevB.81.094525} {\bibfield  {journal} {\bibinfo
  {journal} {Phys. Rev. B}\ }\textbf {\bibinfo {volume} {81}},\ \bibinfo
  {pages} {094525} (\bibinfo {year} {2010})}\BibitemShut {NoStop}%
\bibitem [{\citenamefont {Malashevich}\ and\ \citenamefont
  {Souza}(2010)}]{PhysRevB.82.245118}%
  \BibitemOpen
  \bibfield  {author} {\bibinfo {author} {\bibfnamefont {A.}~\bibnamefont
  {Malashevich}}\ and\ \bibinfo {author} {\bibfnamefont {I.}~\bibnamefont
  {Souza}},\ }\bibfield  {title} {\bibinfo {title} {Band theory of spatial
  dispersion in magnetoelectrics},\ }\href
  {https://doi.org/10.1103/PhysRevB.82.245118} {\bibfield  {journal} {\bibinfo
  {journal} {Phys. Rev. B}\ }\textbf {\bibinfo {volume} {82}},\ \bibinfo
  {pages} {245118} (\bibinfo {year} {2010})}\BibitemShut {NoStop}%
\bibitem [{\citenamefont {Zhong}\ \emph {et~al.}(2016)\citenamefont {Zhong},
  \citenamefont {Moore},\ and\ \citenamefont {Souza}}]{PhysRevLett.116.077201}%
  \BibitemOpen
  \bibfield  {author} {\bibinfo {author} {\bibfnamefont {S.}~\bibnamefont
  {Zhong}}, \bibinfo {author} {\bibfnamefont {J.~E.}\ \bibnamefont {Moore}},\
  and\ \bibinfo {author} {\bibfnamefont {I.}~\bibnamefont {Souza}},\ }\bibfield
   {title} {\bibinfo {title} {Gyrotropic magnetic effect and the magnetic
  moment on the fermi surface},\ }\href
  {https://doi.org/10.1103/PhysRevLett.116.077201} {\bibfield  {journal}
  {\bibinfo  {journal} {Phys. Rev. Lett.}\ }\textbf {\bibinfo {volume} {116}},\
  \bibinfo {pages} {077201} (\bibinfo {year} {2016})}\BibitemShut {NoStop}%
\bibitem [{\citenamefont {Ma}\ and\ \citenamefont
  {Pesin}(2015)}]{PhysRevB.92.235205}%
  \BibitemOpen
  \bibfield  {author} {\bibinfo {author} {\bibfnamefont {J.}~\bibnamefont
  {Ma}}\ and\ \bibinfo {author} {\bibfnamefont {D.~A.}\ \bibnamefont {Pesin}},\
  }\bibfield  {title} {\bibinfo {title} {Chiral magnetic effect and natural
  optical activity in metals with or without weyl points},\ }\href
  {https://doi.org/10.1103/PhysRevB.92.235205} {\bibfield  {journal} {\bibinfo
  {journal} {Phys. Rev. B}\ }\textbf {\bibinfo {volume} {92}},\ \bibinfo
  {pages} {235205} (\bibinfo {year} {2015})}\BibitemShut {NoStop}%
\bibitem [{\citenamefont {Gao}\ and\ \citenamefont
  {Xiao}(2019)}]{PhysRevLett.122.227402}%
  \BibitemOpen
  \bibfield  {author} {\bibinfo {author} {\bibfnamefont {Y.}~\bibnamefont
  {Gao}}\ and\ \bibinfo {author} {\bibfnamefont {D.}~\bibnamefont {Xiao}},\
  }\bibfield  {title} {\bibinfo {title} {Nonreciprocal directional dichroism
  induced by the quantum metric dipole},\ }\href
  {https://doi.org/10.1103/PhysRevLett.122.227402} {\bibfield  {journal}
  {\bibinfo  {journal} {Phys. Rev. Lett.}\ }\textbf {\bibinfo {volume} {122}},\
  \bibinfo {pages} {227402} (\bibinfo {year} {2019})}\BibitemShut {NoStop}%
\bibitem [{\citenamefont {Duff}\ and\ \citenamefont
  {Sipe}(2022)}]{PhysRevB.106.085413}%
  \BibitemOpen
  \bibfield  {author} {\bibinfo {author} {\bibfnamefont {A.~H.}\ \bibnamefont
  {Duff}}\ and\ \bibinfo {author} {\bibfnamefont {J.~E.}\ \bibnamefont
  {Sipe}},\ }\bibfield  {title} {\bibinfo {title} {Magnetoelectric
  polarizability and optical activity: Spin and frequency dependence},\ }\href
  {https://doi.org/10.1103/PhysRevB.106.085413} {\bibfield  {journal} {\bibinfo
   {journal} {Phys. Rev. B}\ }\textbf {\bibinfo {volume} {106}},\ \bibinfo
  {pages} {085413} (\bibinfo {year} {2022})}\BibitemShut {NoStop}%
\bibitem [{\citenamefont {Hidalgo}\ \emph {et~al.}(2009)\citenamefont
  {Hidalgo}, \citenamefont {S\'anchez-Castillo},\ and\ \citenamefont
  {Noguez}}]{PhysRevB.79.075438}%
  \BibitemOpen
  \bibfield  {author} {\bibinfo {author} {\bibfnamefont {F.}~\bibnamefont
  {Hidalgo}}, \bibinfo {author} {\bibfnamefont {A.}~\bibnamefont
  {S\'anchez-Castillo}},\ and\ \bibinfo {author} {\bibfnamefont
  {C.}~\bibnamefont {Noguez}},\ }\bibfield  {title} {\bibinfo {title}
  {Efficient first-principles method for calculating the circular dichroism of
  nanostructures},\ }\href {https://doi.org/10.1103/PhysRevB.79.075438}
  {\bibfield  {journal} {\bibinfo  {journal} {Phys. Rev. B}\ }\textbf {\bibinfo
  {volume} {79}},\ \bibinfo {pages} {075438} (\bibinfo {year}
  {2009})}\BibitemShut {NoStop}%
\bibitem [{\citenamefont {Tsirkin}\ \emph {et~al.}(2018)\citenamefont
  {Tsirkin}, \citenamefont {Puente},\ and\ \citenamefont
  {Souza}}]{PhysRevB.97.035158}%
  \BibitemOpen
  \bibfield  {author} {\bibinfo {author} {\bibfnamefont {S.~S.}\ \bibnamefont
  {Tsirkin}}, \bibinfo {author} {\bibfnamefont {P.~A.}\ \bibnamefont
  {Puente}},\ and\ \bibinfo {author} {\bibfnamefont {I.}~\bibnamefont
  {Souza}},\ }\bibfield  {title} {\bibinfo {title} {Gyrotropic effects in
  trigonal tellurium studied from first principles},\ }\href
  {https://doi.org/10.1103/PhysRevB.97.035158} {\bibfield  {journal} {\bibinfo
  {journal} {Phys. Rev. B}\ }\textbf {\bibinfo {volume} {97}},\ \bibinfo
  {pages} {035158} (\bibinfo {year} {2018})}\BibitemShut {NoStop}%
\bibitem [{\citenamefont {R{\'e}rat}\ and\ \citenamefont
  {Kirtman}(2021)}]{rerat2021first}%
  \BibitemOpen
  \bibfield  {author} {\bibinfo {author} {\bibfnamefont {M.}~\bibnamefont
  {R{\'e}rat}}\ and\ \bibinfo {author} {\bibfnamefont {B.}~\bibnamefont
  {Kirtman}},\ }\bibfield  {title} {\bibinfo {title} {First-principles
  calculation of the optical rotatory power of periodic systems: Application on
  $\alpha$-quartz, tartaric acid crystal, and chiral (n, m)-carbon nanotubes},\
  }\href {https://pubs.acs.org/doi/10.1021/acs.jctc.1c00243} {\bibfield
  {journal} {\bibinfo  {journal} {Journal of Chemical Theory and Computation}\
  }\textbf {\bibinfo {volume} {17}},\ \bibinfo {pages} {4063} (\bibinfo {year}
  {2021})}\BibitemShut {NoStop}%
\bibitem [{\citenamefont {Desmarais}\ \emph {et~al.}(2023)\citenamefont
  {Desmarais}, \citenamefont {Kirtman},\ and\ \citenamefont
  {R\'erat}}]{PhysRevB.107.224430}%
  \BibitemOpen
  \bibfield  {author} {\bibinfo {author} {\bibfnamefont {J.~K.}\ \bibnamefont
  {Desmarais}}, \bibinfo {author} {\bibfnamefont {B.}~\bibnamefont {Kirtman}},\
  and\ \bibinfo {author} {\bibfnamefont {M.}~\bibnamefont {R\'erat}},\
  }\bibfield  {title} {\bibinfo {title} {First-principles calculation of the
  optical rotatory power of periodic systems: Modern theory with modern
  functionals},\ }\href {https://doi.org/10.1103/PhysRevB.107.224430}
  {\bibfield  {journal} {\bibinfo  {journal} {Phys. Rev. B}\ }\textbf {\bibinfo
  {volume} {107}},\ \bibinfo {pages} {224430} (\bibinfo {year}
  {2023})}\BibitemShut {NoStop}%
\bibitem [{\citenamefont {Kim}\ \emph {et~al.}(2016)\citenamefont {Kim},
  \citenamefont {S{\'a}nchez-Castillo}, \citenamefont {Ziegler}, \citenamefont
  {Ogawa}, \citenamefont {Noguez},\ and\ \citenamefont {Park}}]{kim2016chiral}%
  \BibitemOpen
  \bibfield  {author} {\bibinfo {author} {\bibfnamefont {C.-J.}\ \bibnamefont
  {Kim}}, \bibinfo {author} {\bibfnamefont {A.}~\bibnamefont
  {S{\'a}nchez-Castillo}}, \bibinfo {author} {\bibfnamefont {Z.}~\bibnamefont
  {Ziegler}}, \bibinfo {author} {\bibfnamefont {Y.}~\bibnamefont {Ogawa}},
  \bibinfo {author} {\bibfnamefont {C.}~\bibnamefont {Noguez}},\ and\ \bibinfo
  {author} {\bibfnamefont {J.}~\bibnamefont {Park}},\ }\bibfield  {title}
  {\bibinfo {title} {Chiral atomically thin films},\ }\href
  {https://www.nature.com/articles/nnano.2016.3#citeas} {\bibfield  {journal}
  {\bibinfo  {journal} {Nature nanotechnology}\ }\textbf {\bibinfo {volume}
  {11}},\ \bibinfo {pages} {520} (\bibinfo {year} {2016})}\BibitemShut
  {NoStop}%
\bibitem [{\citenamefont {Morell}\ \emph {et~al.}(2017)\citenamefont {Morell},
  \citenamefont {Chico},\ and\ \citenamefont {Brey}}]{SuárezMorell_2017}%
  \BibitemOpen
  \bibfield  {author} {\bibinfo {author} {\bibfnamefont {E.~S.}\ \bibnamefont
  {Morell}}, \bibinfo {author} {\bibfnamefont {L.}~\bibnamefont {Chico}},\ and\
  \bibinfo {author} {\bibfnamefont {L.}~\bibnamefont {Brey}},\ }\bibfield
  {title} {\bibinfo {title} {Twisting dirac fermions: circular dichroism in
  bilayer graphene},\ }\href {https://doi.org/10.1088/2053-1583/aa7eb6}
  {\bibfield  {journal} {\bibinfo  {journal} {2D Materials}\ }\textbf {\bibinfo
  {volume} {4}},\ \bibinfo {pages} {035015} (\bibinfo {year}
  {2017})}\BibitemShut {NoStop}%
\bibitem [{\citenamefont {Stauber}\ \emph {et~al.}(2018)\citenamefont
  {Stauber}, \citenamefont {Low},\ and\ \citenamefont
  {G\'omez-Santos}}]{PhysRevLett.120.046801}%
  \BibitemOpen
  \bibfield  {author} {\bibinfo {author} {\bibfnamefont {T.}~\bibnamefont
  {Stauber}}, \bibinfo {author} {\bibfnamefont {T.}~\bibnamefont {Low}},\ and\
  \bibinfo {author} {\bibfnamefont {G.}~\bibnamefont {G\'omez-Santos}},\
  }\bibfield  {title} {\bibinfo {title} {Chiral response of twisted bilayer
  graphene},\ }\href {https://doi.org/10.1103/PhysRevLett.120.046801}
  {\bibfield  {journal} {\bibinfo  {journal} {Phys. Rev. Lett.}\ }\textbf
  {\bibinfo {volume} {120}},\ \bibinfo {pages} {046801} (\bibinfo {year}
  {2018})}\BibitemShut {NoStop}%
\bibitem [{\citenamefont {Chang}\ \emph {et~al.}(2022)\citenamefont {Chang},
  \citenamefont {Zheng}, \citenamefont {Sipe},\ and\ \citenamefont
  {Cheng}}]{PhysRevB.106.245405}%
  \BibitemOpen
  \bibfield  {author} {\bibinfo {author} {\bibfnamefont {K.}~\bibnamefont
  {Chang}}, \bibinfo {author} {\bibfnamefont {Z.}~\bibnamefont {Zheng}},
  \bibinfo {author} {\bibfnamefont {J.~E.}\ \bibnamefont {Sipe}},\ and\
  \bibinfo {author} {\bibfnamefont {J.~L.}\ \bibnamefont {Cheng}},\ }\bibfield
  {title} {\bibinfo {title} {Theory of optical activity in doped systems with
  application to twisted bilayer graphene},\ }\href
  {https://doi.org/10.1103/PhysRevB.106.245405} {\bibfield  {journal} {\bibinfo
   {journal} {Phys. Rev. B}\ }\textbf {\bibinfo {volume} {106}},\ \bibinfo
  {pages} {245405} (\bibinfo {year} {2022})}\BibitemShut {NoStop}%
\bibitem [{\citenamefont {Han}\ \emph {et~al.}(2023)\citenamefont {Han},
  \citenamefont {Wang}, \citenamefont {Sun}, \citenamefont {Wang},\ and\
  \citenamefont {Tang}}]{https://doi.org/10.1002/adma.202206141}%
  \BibitemOpen
  \bibfield  {author} {\bibinfo {author} {\bibfnamefont {Z.}~\bibnamefont
  {Han}}, \bibinfo {author} {\bibfnamefont {F.}~\bibnamefont {Wang}}, \bibinfo
  {author} {\bibfnamefont {J.}~\bibnamefont {Sun}}, \bibinfo {author}
  {\bibfnamefont {X.}~\bibnamefont {Wang}},\ and\ \bibinfo {author}
  {\bibfnamefont {Z.}~\bibnamefont {Tang}},\ }\bibfield  {title} {\bibinfo
  {title} {Recent advances in ultrathin chiral metasurfaces by twisted
  stacking},\ }\href {https://doi.org/https://doi.org/10.1002/adma.202206141}
  {\bibfield  {journal} {\bibinfo  {journal} {Advanced Materials}\ }\textbf
  {\bibinfo {volume} {35}},\ \bibinfo {pages} {2206141} (\bibinfo {year}
  {2023})}\BibitemShut {NoStop}%
\bibitem [{\citenamefont {Ho}\ and\ \citenamefont
  {Do}(2023)}]{PhysRevB.107.195141}%
  \BibitemOpen
  \bibfield  {author} {\bibinfo {author} {\bibfnamefont {S.~T.}\ \bibnamefont
  {Ho}}\ and\ \bibinfo {author} {\bibfnamefont {V.~N.}\ \bibnamefont {Do}},\
  }\bibfield  {title} {\bibinfo {title} {Optical activity and transport in
  twisted bilayer graphene: Spatial dispersion effects},\ }\href
  {https://doi.org/10.1103/PhysRevB.107.195141} {\bibfield  {journal} {\bibinfo
   {journal} {Phys. Rev. B}\ }\textbf {\bibinfo {volume} {107}},\ \bibinfo
  {pages} {195141} (\bibinfo {year} {2023})}\BibitemShut {NoStop}%
\bibitem [{\citenamefont {Ahn}\ \emph {et~al.}(2022)\citenamefont {Ahn},
  \citenamefont {Xu},\ and\ \citenamefont {Vishwanath}}]{ahn2022theory}%
  \BibitemOpen
  \bibfield  {author} {\bibinfo {author} {\bibfnamefont {J.}~\bibnamefont
  {Ahn}}, \bibinfo {author} {\bibfnamefont {S.-Y.}\ \bibnamefont {Xu}},\ and\
  \bibinfo {author} {\bibfnamefont {A.}~\bibnamefont {Vishwanath}},\ }\bibfield
   {title} {\bibinfo {title} {Theory of optical axion electrodynamics and
  application to the kerr effect in topological antiferromagnets},\ }\href
  {https://www.nature.com/articles/s41467-022-35248-8} {\bibfield  {journal}
  {\bibinfo  {journal} {Nature Communications}\ }\textbf {\bibinfo {volume}
  {13}},\ \bibinfo {pages} {7615} (\bibinfo {year} {2022})}\BibitemShut
  {NoStop}%
\bibitem [{\citenamefont {Óscar Pozo~Ocaña}\ and\ \citenamefont
  {Souza}(2023)}]{10.21468/SciPostPhys.14.5.118}%
  \BibitemOpen
  \bibfield  {author} {\bibinfo {author} {\bibnamefont {Óscar Pozo~Ocaña}}\
  and\ \bibinfo {author} {\bibfnamefont {I.}~\bibnamefont {Souza}},\ }\bibfield
   {title} {\bibinfo {title} {{Multipole theory of optical spatial dispersion
  in crystals}},\ }\href {https://doi.org/10.21468/SciPostPhys.14.5.118}
  {\bibfield  {journal} {\bibinfo  {journal} {SciPost Phys.}\ }\textbf
  {\bibinfo {volume} {14}},\ \bibinfo {pages} {118} (\bibinfo {year}
  {2023})}\BibitemShut {NoStop}%
\bibitem [{\citenamefont {Wang}\ and\ \citenamefont
  {Yan}(2023)}]{PhysRevB.107.045201}%
  \BibitemOpen
  \bibfield  {author} {\bibinfo {author} {\bibfnamefont {X.}~\bibnamefont
  {Wang}}\ and\ \bibinfo {author} {\bibfnamefont {Y.}~\bibnamefont {Yan}},\
  }\bibfield  {title} {\bibinfo {title} {Optical activity of solids from first
  principles},\ }\href {https://doi.org/10.1103/PhysRevB.107.045201} {\bibfield
   {journal} {\bibinfo  {journal} {Phys. Rev. B}\ }\textbf {\bibinfo {volume}
  {107}},\ \bibinfo {pages} {045201} (\bibinfo {year} {2023})}\BibitemShut
  {NoStop}%
\bibitem [{\citenamefont {Mineev}(2013)}]{PhysRevB.88.134514}%
  \BibitemOpen
  \bibfield  {author} {\bibinfo {author} {\bibfnamefont {V.~P.}\ \bibnamefont
  {Mineev}},\ }\bibfield  {title} {\bibinfo {title} {Magnetostatics and optics
  of noncentrosymmetric metals},\ }\href
  {https://doi.org/10.1103/PhysRevB.88.134514} {\bibfield  {journal} {\bibinfo
  {journal} {Phys. Rev. B}\ }\textbf {\bibinfo {volume} {88}},\ \bibinfo
  {pages} {134514} (\bibinfo {year} {2013})}\BibitemShut {NoStop}%
\bibitem [{\citenamefont {Bauer}\ and\ \citenamefont
  {Sigrist}(2012)}]{bauer2012non}%
  \BibitemOpen
  \bibfield  {author} {\bibinfo {author} {\bibfnamefont {E.}~\bibnamefont
  {Bauer}}\ and\ \bibinfo {author} {\bibfnamefont {M.}~\bibnamefont
  {Sigrist}},\ }\href@noop {} {\emph {\bibinfo {title} {Non-centrosymmetric
  superconductors: introduction and overview}}},\ Vol.\ \bibinfo {volume}
  {847}\ (\bibinfo  {publisher} {Springer Science \& Business Media},\ \bibinfo
  {year} {2012})\BibitemShut {NoStop}%
\bibitem [{\citenamefont {Edelstein}(1995)}]{PhysRevLett.75.2004}%
  \BibitemOpen
  \bibfield  {author} {\bibinfo {author} {\bibfnamefont {V.~M.}\ \bibnamefont
  {Edelstein}},\ }\bibfield  {title} {\bibinfo {title} {Magnetoelectric effect
  in polar superconductors},\ }\href
  {https://doi.org/10.1103/PhysRevLett.75.2004} {\bibfield  {journal} {\bibinfo
   {journal} {Phys. Rev. Lett.}\ }\textbf {\bibinfo {volume} {75}},\ \bibinfo
  {pages} {2004} (\bibinfo {year} {1995})}\BibitemShut {NoStop}%
\bibitem [{\citenamefont {Wakatsuki}\ and\ \citenamefont
  {Nagaosa}(2018)}]{PhysRevLett.121.026601}%
  \BibitemOpen
  \bibfield  {author} {\bibinfo {author} {\bibfnamefont {R.}~\bibnamefont
  {Wakatsuki}}\ and\ \bibinfo {author} {\bibfnamefont {N.}~\bibnamefont
  {Nagaosa}},\ }\bibfield  {title} {\bibinfo {title} {Nonreciprocal current in
  noncentrosymmetric rashba superconductors},\ }\href
  {https://doi.org/10.1103/PhysRevLett.121.026601} {\bibfield  {journal}
  {\bibinfo  {journal} {Phys. Rev. Lett.}\ }\textbf {\bibinfo {volume} {121}},\
  \bibinfo {pages} {026601} (\bibinfo {year} {2018})}\BibitemShut {NoStop}%
\bibitem [{\citenamefont {Wakatsuki}\ \emph {et~al.}(2017)\citenamefont
  {Wakatsuki}, \citenamefont {Saito}, \citenamefont {Hoshino}, \citenamefont
  {Itahashi}, \citenamefont {Ideue}, \citenamefont {Ezawa}, \citenamefont
  {Iwasa},\ and\ \citenamefont {Nagaosa}}]{doi:10.1126/sciadv.1602390}%
  \BibitemOpen
  \bibfield  {author} {\bibinfo {author} {\bibfnamefont {R.}~\bibnamefont
  {Wakatsuki}}, \bibinfo {author} {\bibfnamefont {Y.}~\bibnamefont {Saito}},
  \bibinfo {author} {\bibfnamefont {S.}~\bibnamefont {Hoshino}}, \bibinfo
  {author} {\bibfnamefont {Y.~M.}\ \bibnamefont {Itahashi}}, \bibinfo {author}
  {\bibfnamefont {T.}~\bibnamefont {Ideue}}, \bibinfo {author} {\bibfnamefont
  {M.}~\bibnamefont {Ezawa}}, \bibinfo {author} {\bibfnamefont
  {Y.}~\bibnamefont {Iwasa}},\ and\ \bibinfo {author} {\bibfnamefont
  {N.}~\bibnamefont {Nagaosa}},\ }\bibfield  {title} {\bibinfo {title}
  {Nonreciprocal charge transport in noncentrosymmetric superconductors},\
  }\href {https://doi.org/10.1126/sciadv.1602390} {\bibfield  {journal}
  {\bibinfo  {journal} {Science Advances}\ }\textbf {\bibinfo {volume} {3}},\
  \bibinfo {pages} {e1602390} (\bibinfo {year} {2017})}\BibitemShut {NoStop}%
\bibitem [{\citenamefont {Ideue}\ \emph {et~al.}(2020)\citenamefont {Ideue},
  \citenamefont {Koshikawa}, \citenamefont {Namiki}, \citenamefont {Sasagawa},\
  and\ \citenamefont {Iwasa}}]{PhysRevResearch.2.042046}%
  \BibitemOpen
  \bibfield  {author} {\bibinfo {author} {\bibfnamefont {T.}~\bibnamefont
  {Ideue}}, \bibinfo {author} {\bibfnamefont {S.}~\bibnamefont {Koshikawa}},
  \bibinfo {author} {\bibfnamefont {H.}~\bibnamefont {Namiki}}, \bibinfo
  {author} {\bibfnamefont {T.}~\bibnamefont {Sasagawa}},\ and\ \bibinfo
  {author} {\bibfnamefont {Y.}~\bibnamefont {Iwasa}},\ }\bibfield  {title}
  {\bibinfo {title} {Giant nonreciprocal magnetotransport in bulk trigonal
  superconductor $\mathrm{PbTa}{\mathrm{se}}_{2}$},\ }\href
  {https://doi.org/10.1103/PhysRevResearch.2.042046} {\bibfield  {journal}
  {\bibinfo  {journal} {Phys. Rev. Res.}\ }\textbf {\bibinfo {volume} {2}},\
  \bibinfo {pages} {042046} (\bibinfo {year} {2020})}\BibitemShut {NoStop}%
\bibitem [{\citenamefont {Daido}\ \emph {et~al.}(2022)\citenamefont {Daido},
  \citenamefont {Ikeda},\ and\ \citenamefont
  {Yanase}}]{PhysRevLett.128.037001}%
  \BibitemOpen
  \bibfield  {author} {\bibinfo {author} {\bibfnamefont {A.}~\bibnamefont
  {Daido}}, \bibinfo {author} {\bibfnamefont {Y.}~\bibnamefont {Ikeda}},\ and\
  \bibinfo {author} {\bibfnamefont {Y.}~\bibnamefont {Yanase}},\ }\bibfield
  {title} {\bibinfo {title} {Intrinsic superconducting diode effect},\ }\href
  {https://doi.org/10.1103/PhysRevLett.128.037001} {\bibfield  {journal}
  {\bibinfo  {journal} {Phys. Rev. Lett.}\ }\textbf {\bibinfo {volume} {128}},\
  \bibinfo {pages} {037001} (\bibinfo {year} {2022})}\BibitemShut {NoStop}%
\bibitem [{\citenamefont {Yuan}\ and\ \citenamefont
  {Fu}(2022)}]{doi:10.1073/pnas.2119548119}%
  \BibitemOpen
  \bibfield  {author} {\bibinfo {author} {\bibfnamefont {N.~F.~Q.}\
  \bibnamefont {Yuan}}\ and\ \bibinfo {author} {\bibfnamefont {L.}~\bibnamefont
  {Fu}},\ }\bibfield  {title} {\bibinfo {title} {Supercurrent diode effect and
  finite-momentum superconductors},\ }\href
  {https://doi.org/10.1073/pnas.2119548119} {\bibfield  {journal} {\bibinfo
  {journal} {Proceedings of the National Academy of Sciences}\ }\textbf
  {\bibinfo {volume} {119}},\ \bibinfo {pages} {e2119548119} (\bibinfo {year}
  {2022})}\BibitemShut {NoStop}%
\bibitem [{\citenamefont {He}\ \emph {et~al.}(2022)\citenamefont {He},
  \citenamefont {Tanaka},\ and\ \citenamefont {Nagaosa}}]{He_2022}%
  \BibitemOpen
  \bibfield  {author} {\bibinfo {author} {\bibfnamefont {J.~J.}\ \bibnamefont
  {He}}, \bibinfo {author} {\bibfnamefont {Y.}~\bibnamefont {Tanaka}},\ and\
  \bibinfo {author} {\bibfnamefont {N.}~\bibnamefont {Nagaosa}},\ }\bibfield
  {title} {\bibinfo {title} {A phenomenological theory of superconductor
  diodes},\ }\href {https://doi.org/10.1088/1367-2630/ac6766} {\bibfield
  {journal} {\bibinfo  {journal} {New Journal of Physics}\ }\textbf {\bibinfo
  {volume} {24}},\ \bibinfo {pages} {053014} (\bibinfo {year}
  {2022})}\BibitemShut {NoStop}%
\bibitem [{\citenamefont {Ando}\ \emph {et~al.}(2020)\citenamefont {Ando},
  \citenamefont {Miyasaka}, \citenamefont {Li}, \citenamefont {Ishizuka},
  \citenamefont {Arakawa}, \citenamefont {Shiota}, \citenamefont {Moriyama},
  \citenamefont {Yanase},\ and\ \citenamefont {Ono}}]{ando2020observation}%
  \BibitemOpen
  \bibfield  {author} {\bibinfo {author} {\bibfnamefont {F.}~\bibnamefont
  {Ando}}, \bibinfo {author} {\bibfnamefont {Y.}~\bibnamefont {Miyasaka}},
  \bibinfo {author} {\bibfnamefont {T.}~\bibnamefont {Li}}, \bibinfo {author}
  {\bibfnamefont {J.}~\bibnamefont {Ishizuka}}, \bibinfo {author}
  {\bibfnamefont {T.}~\bibnamefont {Arakawa}}, \bibinfo {author} {\bibfnamefont
  {Y.}~\bibnamefont {Shiota}}, \bibinfo {author} {\bibfnamefont
  {T.}~\bibnamefont {Moriyama}}, \bibinfo {author} {\bibfnamefont
  {Y.}~\bibnamefont {Yanase}},\ and\ \bibinfo {author} {\bibfnamefont
  {T.}~\bibnamefont {Ono}},\ }\bibfield  {title} {\bibinfo {title} {Observation
  of superconducting diode effect},\ }\href
  {https://www.nature.com/articles/s41586-020-2590-4} {\bibfield  {journal}
  {\bibinfo  {journal} {Nature}\ }\textbf {\bibinfo {volume} {584}},\ \bibinfo
  {pages} {373} (\bibinfo {year} {2020})}\BibitemShut {NoStop}%
\bibitem [{\citenamefont {Kubo}(1957)}]{doi:10.1143/JPSJ.12.570}%
  \BibitemOpen
  \bibfield  {author} {\bibinfo {author} {\bibfnamefont {R.}~\bibnamefont
  {Kubo}},\ }\bibfield  {title} {\bibinfo {title} {Statistical-mechanical
  theory of irreversible processes. i. general theory and simple applications
  to magnetic and conduction problems},\ }\href
  {https://doi.org/10.1143/JPSJ.12.570} {\bibfield  {journal} {\bibinfo
  {journal} {Journal of the Physical Society of Japan}\ }\textbf {\bibinfo
  {volume} {12}},\ \bibinfo {pages} {570} (\bibinfo {year} {1957})}\BibitemShut
  {NoStop}%
\bibitem [{\citenamefont {Nomura}(1960)}]{PhysRevLett.5.500}%
  \BibitemOpen
  \bibfield  {author} {\bibinfo {author} {\bibfnamefont {K.~C.}\ \bibnamefont
  {Nomura}},\ }\bibfield  {title} {\bibinfo {title} {Optical activity in
  tellurium},\ }\href {https://doi.org/10.1103/PhysRevLett.5.500} {\bibfield
  {journal} {\bibinfo  {journal} {Phys. Rev. Lett.}\ }\textbf {\bibinfo
  {volume} {5}},\ \bibinfo {pages} {500} (\bibinfo {year} {1960})}\BibitemShut
  {NoStop}%
\bibitem [{\citenamefont {Hayami}\ \emph {et~al.}(2018)\citenamefont {Hayami},
  \citenamefont {Yatsushiro}, \citenamefont {Yanagi},\ and\ \citenamefont
  {Kusunose}}]{PhysRevB.98.165110}%
  \BibitemOpen
  \bibfield  {author} {\bibinfo {author} {\bibfnamefont {S.}~\bibnamefont
  {Hayami}}, \bibinfo {author} {\bibfnamefont {M.}~\bibnamefont {Yatsushiro}},
  \bibinfo {author} {\bibfnamefont {Y.}~\bibnamefont {Yanagi}},\ and\ \bibinfo
  {author} {\bibfnamefont {H.}~\bibnamefont {Kusunose}},\ }\bibfield  {title}
  {\bibinfo {title} {Classification of atomic-scale multipoles under
  crystallographic point groups and application to linear response tensors},\
  }\href {https://doi.org/10.1103/PhysRevB.98.165110} {\bibfield  {journal}
  {\bibinfo  {journal} {Phys. Rev. B}\ }\textbf {\bibinfo {volume} {98}},\
  \bibinfo {pages} {165110} (\bibinfo {year} {2018})}\BibitemShut {NoStop}%
\bibitem [{\citenamefont {Kishine}\ \emph {et~al.}(2022)\citenamefont
  {Kishine}, \citenamefont {Kusunose},\ and\ \citenamefont
  {Yamamoto}}]{doi.org/10.1002/ijch.202200049}%
  \BibitemOpen
  \bibfield  {author} {\bibinfo {author} {\bibfnamefont {J.-i.}\ \bibnamefont
  {Kishine}}, \bibinfo {author} {\bibfnamefont {H.}~\bibnamefont {Kusunose}},\
  and\ \bibinfo {author} {\bibfnamefont {H.~M.}\ \bibnamefont {Yamamoto}},\
  }\bibfield  {title} {\bibinfo {title} {On the definition of chirality and
  enantioselective fields},\ }\href
  {https://doi.org/https://doi.org/10.1002/ijch.202200049} {\bibfield
  {journal} {\bibinfo  {journal} {Israel Journal of Chemistry}\ }\textbf
  {\bibinfo {volume} {62}},\ \bibinfo {pages} {e202200049} (\bibinfo {year}
  {2022})}\BibitemShut {NoStop}%
\bibitem [{\citenamefont {Jerphagnon}\ and\ \citenamefont
  {Chemla}(2008)}]{10.1063/1.433207}%
  \BibitemOpen
  \bibfield  {author} {\bibinfo {author} {\bibfnamefont {J.}~\bibnamefont
  {Jerphagnon}}\ and\ \bibinfo {author} {\bibfnamefont {D.~S.}\ \bibnamefont
  {Chemla}},\ }\bibfield  {title} {\bibinfo {title} {{Optical activity of
  crystals}},\ }\href {https://doi.org/10.1063/1.433207} {\bibfield  {journal}
  {\bibinfo  {journal} {The Journal of Chemical Physics}\ }\textbf {\bibinfo
  {volume} {65}},\ \bibinfo {pages} {1522} (\bibinfo {year}
  {2008})}\BibitemShut {NoStop}%
\bibitem [{\citenamefont {Graham}\ and\ \citenamefont
  {Raab}(1996)}]{Graham:96}%
  \BibitemOpen
  \bibfield  {author} {\bibinfo {author} {\bibfnamefont {E.~B.}\ \bibnamefont
  {Graham}}\ and\ \bibinfo {author} {\bibfnamefont {R.~E.}\ \bibnamefont
  {Raab}},\ }\bibfield  {title} {\bibinfo {title} {Reflection from
  noncentrosymmetric uniaxial crystals: a multipole approach},\ }\href
  {https://doi.org/10.1364/JOSAA.13.001239} {\bibfield  {journal} {\bibinfo
  {journal} {J. Opt. Soc. Am. A}\ }\textbf {\bibinfo {volume} {13}},\ \bibinfo
  {pages} {1239} (\bibinfo {year} {1996})}\BibitemShut {NoStop}%
\bibitem [{\citenamefont {Norman}(2015{\natexlab{a}})}]{PhysRevB.92.241116}%
  \BibitemOpen
  \bibfield  {author} {\bibinfo {author} {\bibfnamefont {M.~R.}\ \bibnamefont
  {Norman}},\ }\bibfield  {title} {\bibinfo {title} {Vector optical activity in
  the weyl semimetal taas},\ }\href
  {https://doi.org/10.1103/PhysRevB.92.241116} {\bibfield  {journal} {\bibinfo
  {journal} {Phys. Rev. B}\ }\textbf {\bibinfo {volume} {92}},\ \bibinfo
  {pages} {241116} (\bibinfo {year} {2015}{\natexlab{a}})}\BibitemShut
  {NoStop}%
\bibitem [{\citenamefont {Ivchenko}\ \emph
  {et~al.}(1978{\natexlab{a}})\citenamefont {Ivchenko}, \citenamefont
  {Permogorov},\ and\ \citenamefont {Selkin}}]{Ivchenko1978}%
  \BibitemOpen
  \bibfield  {author} {\bibinfo {author} {\bibfnamefont {E.~L.}\ \bibnamefont
  {Ivchenko}}, \bibinfo {author} {\bibfnamefont {S.~A.}\ \bibnamefont
  {Permogorov}},\ and\ \bibinfo {author} {\bibfnamefont {A.~V.}\ \bibnamefont
  {Selkin}},\ }\bibfield  {title} {\bibinfo {title} {Natural optical activity
  of {CdS} crystals in the exciton region of the spectrum},\ }\href@noop {}
  {\bibfield  {journal} {\bibinfo  {journal} {Pis’ma Zh. Eksp. Teor. Fiz.}\
  }\textbf {\bibinfo {volume} {27}},\ \bibinfo {pages} {27} (\bibinfo {year}
  {1978}{\natexlab{a}})},\ \bibinfo {note} {[JETP Lett. 27, 24
  (1978)]}\BibitemShut {NoStop}%
\bibitem [{\citenamefont {Ivchenko}\ \emph
  {et~al.}(1978{\natexlab{b}})\citenamefont {Ivchenko}, \citenamefont
  {Permogorov},\ and\ \citenamefont {Sel'kin}}]{IVCHENKO1978345}%
  \BibitemOpen
  \bibfield  {author} {\bibinfo {author} {\bibfnamefont {E.}~\bibnamefont
  {Ivchenko}}, \bibinfo {author} {\bibfnamefont {S.}~\bibnamefont
  {Permogorov}},\ and\ \bibinfo {author} {\bibfnamefont {A.}~\bibnamefont
  {Sel'kin}},\ }\bibfield  {title} {\bibinfo {title} {Optical activity of cds
  crystals in exciton spectral region},\ }\href
  {https://doi.org/https://doi.org/10.1016/0038-1098(78)90438-6} {\bibfield
  {journal} {\bibinfo  {journal} {Solid State Communications}\ }\textbf
  {\bibinfo {volume} {28}},\ \bibinfo {pages} {345} (\bibinfo {year}
  {1978}{\natexlab{b}})}\BibitemShut {NoStop}%
\bibitem [{\citenamefont {Norman}(2015{\natexlab{b}})}]{PhysRevB.92.075113}%
  \BibitemOpen
  \bibfield  {author} {\bibinfo {author} {\bibfnamefont {M.~R.}\ \bibnamefont
  {Norman}},\ }\bibfield  {title} {\bibinfo {title} {Dichroism as a probe for
  parity-breaking phases of spin-orbit coupled metals},\ }\href
  {https://doi.org/10.1103/PhysRevB.92.075113} {\bibfield  {journal} {\bibinfo
  {journal} {Phys. Rev. B}\ }\textbf {\bibinfo {volume} {92}},\ \bibinfo
  {pages} {075113} (\bibinfo {year} {2015}{\natexlab{b}})}\BibitemShut
  {NoStop}%
\bibitem [{\citenamefont {Shibata}\ \emph {et~al.}(2016)\citenamefont
  {Shibata}, \citenamefont {Takeuchi}, \citenamefont {Kohno},\ and\
  \citenamefont {Tatara}}]{doi:10.7566/JPSJ.85.033701}%
  \BibitemOpen
  \bibfield  {author} {\bibinfo {author} {\bibfnamefont {J.}~\bibnamefont
  {Shibata}}, \bibinfo {author} {\bibfnamefont {A.}~\bibnamefont {Takeuchi}},
  \bibinfo {author} {\bibfnamefont {H.}~\bibnamefont {Kohno}},\ and\ \bibinfo
  {author} {\bibfnamefont {G.}~\bibnamefont {Tatara}},\ }\bibfield  {title}
  {\bibinfo {title} {Theory of anomalous optical properties of bulk rashba
  conductor},\ }\href {https://doi.org/10.7566/JPSJ.85.033701} {\bibfield
  {journal} {\bibinfo  {journal} {Journal of the Physical Society of Japan}\
  }\textbf {\bibinfo {volume} {85}},\ \bibinfo {pages} {033701} (\bibinfo
  {year} {2016})}\BibitemShut {NoStop}%
\bibitem [{\citenamefont {Pisarev}\ \emph {et~al.}(1991)\citenamefont
  {Pisarev}, \citenamefont {Krichevtsov},\ and\ \citenamefont
  {Pavlov}}]{doi:10.1080/01411599108203448}%
  \BibitemOpen
  \bibfield  {author} {\bibinfo {author} {\bibfnamefont {R.~V.}\ \bibnamefont
  {Pisarev}}, \bibinfo {author} {\bibfnamefont {B.~B.}\ \bibnamefont
  {Krichevtsov}},\ and\ \bibinfo {author} {\bibfnamefont {V.~V.}\ \bibnamefont
  {Pavlov}},\ }\bibfield  {title} {\bibinfo {title} {Optical study of the
  antiferromagnetic-paramagnetic phase transition in chromium oxide
  $\mathrm{Cr_2O_3}$},\ }\href {https://doi.org/10.1080/01411599108203448}
  {\bibfield  {journal} {\bibinfo  {journal} {Phase Transitions}\ }\textbf
  {\bibinfo {volume} {37}},\ \bibinfo {pages} {63} (\bibinfo {year}
  {1991})}\BibitemShut {NoStop}%
\bibitem [{\citenamefont {Krichevtsov}\ \emph {et~al.}(1993)\citenamefont
  {Krichevtsov}, \citenamefont {Pavlov}, \citenamefont {Pisarev},\ and\
  \citenamefont {Gridnev}}]{Krichevtsov_1993}%
  \BibitemOpen
  \bibfield  {author} {\bibinfo {author} {\bibfnamefont {B.~B.}\ \bibnamefont
  {Krichevtsov}}, \bibinfo {author} {\bibfnamefont {V.~V.}\ \bibnamefont
  {Pavlov}}, \bibinfo {author} {\bibfnamefont {R.~V.}\ \bibnamefont
  {Pisarev}},\ and\ \bibinfo {author} {\bibfnamefont {V.~N.}\ \bibnamefont
  {Gridnev}},\ }\bibfield  {title} {\bibinfo {title} {Spontaneous
  non-reciprocal reflection of light from antiferromagnetic cr2o3},\ }\href
  {https://doi.org/10.1088/0953-8984/5/44/014} {\bibfield  {journal} {\bibinfo
  {journal} {Journal of Physics: Condensed Matter}\ }\textbf {\bibinfo {volume}
  {5}},\ \bibinfo {pages} {8233} (\bibinfo {year} {1993})}\BibitemShut
  {NoStop}%
\bibitem [{\citenamefont {Kimura}\ \emph {et~al.}(2020)\citenamefont {Kimura},
  \citenamefont {Katsuyoshi}, \citenamefont {Sawada}, \citenamefont {Kimura},\
  and\ \citenamefont {Kimura}}]{kimura2020imaging}%
  \BibitemOpen
  \bibfield  {author} {\bibinfo {author} {\bibfnamefont {K.}~\bibnamefont
  {Kimura}}, \bibinfo {author} {\bibfnamefont {T.}~\bibnamefont {Katsuyoshi}},
  \bibinfo {author} {\bibfnamefont {Y.}~\bibnamefont {Sawada}}, \bibinfo
  {author} {\bibfnamefont {S.}~\bibnamefont {Kimura}},\ and\ \bibinfo {author}
  {\bibfnamefont {T.}~\bibnamefont {Kimura}},\ }\bibfield  {title} {\bibinfo
  {title} {Imaging switchable magnetoelectric quadrupole domains via
  nonreciprocal linear dichroism},\ }\href
  {https://www.nature.com/articles/s43246-020-0040-3#citeas} {\bibfield
  {journal} {\bibinfo  {journal} {Communications Materials}\ }\textbf {\bibinfo
  {volume} {1}},\ \bibinfo {pages} {39} (\bibinfo {year} {2020})}\BibitemShut
  {NoStop}%
\bibitem [{\citenamefont {Sato}\ \emph {et~al.}(2022)\citenamefont {Sato},
  \citenamefont {Abe}, \citenamefont {Tokunaga},\ and\ \citenamefont
  {Arima}}]{PhysRevB.105.094417}%
  \BibitemOpen
  \bibfield  {author} {\bibinfo {author} {\bibfnamefont {T.}~\bibnamefont
  {Sato}}, \bibinfo {author} {\bibfnamefont {N.}~\bibnamefont {Abe}}, \bibinfo
  {author} {\bibfnamefont {Y.}~\bibnamefont {Tokunaga}},\ and\ \bibinfo
  {author} {\bibfnamefont {T.-h.}\ \bibnamefont {Arima}},\ }\bibfield  {title}
  {\bibinfo {title} {Antiferromagnetic domain wall dynamics in magnetoelectric
  $\mathrm{MnTiO}_{3}$ studied by optical imaging},\ }\href
  {https://doi.org/10.1103/PhysRevB.105.094417} {\bibfield  {journal} {\bibinfo
   {journal} {Phys. Rev. B}\ }\textbf {\bibinfo {volume} {105}},\ \bibinfo
  {pages} {094417} (\bibinfo {year} {2022})}\BibitemShut {NoStop}%
\bibitem [{\citenamefont {Hayashida}\ \emph {et~al.}(2022)\citenamefont
  {Hayashida}, \citenamefont {Arakawa}, \citenamefont {Oshima}, \citenamefont
  {Kimura},\ and\ \citenamefont {Kimura}}]{PhysRevResearch.4.043063}%
  \BibitemOpen
  \bibfield  {author} {\bibinfo {author} {\bibfnamefont {T.}~\bibnamefont
  {Hayashida}}, \bibinfo {author} {\bibfnamefont {K.}~\bibnamefont {Arakawa}},
  \bibinfo {author} {\bibfnamefont {T.}~\bibnamefont {Oshima}}, \bibinfo
  {author} {\bibfnamefont {K.}~\bibnamefont {Kimura}},\ and\ \bibinfo {author}
  {\bibfnamefont {T.}~\bibnamefont {Kimura}},\ }\bibfield  {title} {\bibinfo
  {title} {Observation of antiferromagnetic domains in
  $\mathrm{Cr}_{2}\mathrm{O}_{3}$ using nonreciprocal optical effects},\ }\href
  {https://doi.org/10.1103/PhysRevResearch.4.043063} {\bibfield  {journal}
  {\bibinfo  {journal} {Phys. Rev. Res.}\ }\textbf {\bibinfo {volume} {4}},\
  \bibinfo {pages} {043063} (\bibinfo {year} {2022})}\BibitemShut {NoStop}%
\bibitem [{\citenamefont {K\'ezsm\'arki}\ \emph {et~al.}(2011)\citenamefont
  {K\'ezsm\'arki}, \citenamefont {Kida}, \citenamefont {Murakawa},
  \citenamefont {Bord\'acs}, \citenamefont {Onose},\ and\ \citenamefont
  {Tokura}}]{PhysRevLett.106.057403}%
  \BibitemOpen
  \bibfield  {author} {\bibinfo {author} {\bibfnamefont {I.}~\bibnamefont
  {K\'ezsm\'arki}}, \bibinfo {author} {\bibfnamefont {N.}~\bibnamefont {Kida}},
  \bibinfo {author} {\bibfnamefont {H.}~\bibnamefont {Murakawa}}, \bibinfo
  {author} {\bibfnamefont {S.}~\bibnamefont {Bord\'acs}}, \bibinfo {author}
  {\bibfnamefont {Y.}~\bibnamefont {Onose}},\ and\ \bibinfo {author}
  {\bibfnamefont {Y.}~\bibnamefont {Tokura}},\ }\bibfield  {title} {\bibinfo
  {title} {Enhanced directional dichroism of terahertz light in resonance with
  magnetic excitations of the multiferroic
  ${\mathrm{ba}}_{2}{\mathrm{coge}}_{2}{\mathrm{o}}_{7}$ oxide compound},\
  }\href {https://doi.org/10.1103/PhysRevLett.106.057403} {\bibfield  {journal}
  {\bibinfo  {journal} {Phys. Rev. Lett.}\ }\textbf {\bibinfo {volume} {106}},\
  \bibinfo {pages} {057403} (\bibinfo {year} {2011})}\BibitemShut {NoStop}%
\bibitem [{\citenamefont {Takahashi}\ \emph {et~al.}(2012)\citenamefont
  {Takahashi}, \citenamefont {Shimano}, \citenamefont {Kaneko}, \citenamefont
  {Murakawa},\ and\ \citenamefont {Tokura}}]{takahashi2012magnetoelectric}%
  \BibitemOpen
  \bibfield  {author} {\bibinfo {author} {\bibfnamefont {Y.}~\bibnamefont
  {Takahashi}}, \bibinfo {author} {\bibfnamefont {R.}~\bibnamefont {Shimano}},
  \bibinfo {author} {\bibfnamefont {Y.}~\bibnamefont {Kaneko}}, \bibinfo
  {author} {\bibfnamefont {H.}~\bibnamefont {Murakawa}},\ and\ \bibinfo
  {author} {\bibfnamefont {Y.}~\bibnamefont {Tokura}},\ }\bibfield  {title}
  {\bibinfo {title} {Magnetoelectric resonance with electromagnons in a
  perovskite helimagnet},\ }\href
  {https://www.nature.com/articles/nphys2161#further-reading} {\bibfield
  {journal} {\bibinfo  {journal} {Nature Physics}\ }\textbf {\bibinfo {volume}
  {8}},\ \bibinfo {pages} {121} (\bibinfo {year} {2012})}\BibitemShut {NoStop}%
\bibitem [{\citenamefont {He}\ and\ \citenamefont
  {Law}(2021)}]{PhysRevResearch.3.L032012}%
  \BibitemOpen
  \bibfield  {author} {\bibinfo {author} {\bibfnamefont {W.-Y.}\ \bibnamefont
  {He}}\ and\ \bibinfo {author} {\bibfnamefont {K.~T.}\ \bibnamefont {Law}},\
  }\bibfield  {title} {\bibinfo {title} {Superconducting orbital
  magnetoelectric effect and its evolution across the superconductor-normal
  metal phase transition},\ }\href
  {https://doi.org/10.1103/PhysRevResearch.3.L032012} {\bibfield  {journal}
  {\bibinfo  {journal} {Phys. Rev. Res.}\ }\textbf {\bibinfo {volume} {3}},\
  \bibinfo {pages} {L032012} (\bibinfo {year} {2021})}\BibitemShut {NoStop}%
\bibitem [{\citenamefont {Tinkham}(2004)}]{tinkham2004introduction}%
  \BibitemOpen
  \bibfield  {author} {\bibinfo {author} {\bibfnamefont {M.}~\bibnamefont
  {Tinkham}},\ }\href@noop {} {\emph {\bibinfo {title} {Introduction to
  superconductivity}}}\ (\bibinfo  {publisher} {Courier Corporation},\ \bibinfo
  {year} {2004})\BibitemShut {NoStop}%
\bibitem [{\citenamefont {Watanabe}\ \emph
  {et~al.}(2022{\natexlab{b}})\citenamefont {Watanabe}, \citenamefont {Daido},\
  and\ \citenamefont {Yanase}}]{PhysRevB.105.L100504}%
  \BibitemOpen
  \bibfield  {author} {\bibinfo {author} {\bibfnamefont {H.}~\bibnamefont
  {Watanabe}}, \bibinfo {author} {\bibfnamefont {A.}~\bibnamefont {Daido}},\
  and\ \bibinfo {author} {\bibfnamefont {Y.}~\bibnamefont {Yanase}},\
  }\bibfield  {title} {\bibinfo {title} {Nonreciprocal meissner response in
  parity-mixed superconductors},\ }\href
  {https://doi.org/10.1103/PhysRevB.105.L100504} {\bibfield  {journal}
  {\bibinfo  {journal} {Phys. Rev. B}\ }\textbf {\bibinfo {volume} {105}},\
  \bibinfo {pages} {L100504} (\bibinfo {year}
  {2022}{\natexlab{b}})}\BibitemShut {NoStop}%
\bibitem [{\citenamefont {Gor'kov}\ and\ \citenamefont
  {Rashba}(2001)}]{PhysRevLett.87.037004}%
  \BibitemOpen
  \bibfield  {author} {\bibinfo {author} {\bibfnamefont {L.~P.}\ \bibnamefont
  {Gor'kov}}\ and\ \bibinfo {author} {\bibfnamefont {E.~I.}\ \bibnamefont
  {Rashba}},\ }\bibfield  {title} {\bibinfo {title} {Superconducting 2d system
  with lifted spin degeneracy: Mixed singlet-triplet state},\ }\href
  {https://doi.org/10.1103/PhysRevLett.87.037004} {\bibfield  {journal}
  {\bibinfo  {journal} {Phys. Rev. Lett.}\ }\textbf {\bibinfo {volume} {87}},\
  \bibinfo {pages} {037004} (\bibinfo {year} {2001})}\BibitemShut {NoStop}%
\bibitem [{\citenamefont {Yip}(2002)}]{PhysRevB.65.144508}%
  \BibitemOpen
  \bibfield  {author} {\bibinfo {author} {\bibfnamefont {S.~K.}\ \bibnamefont
  {Yip}},\ }\bibfield  {title} {\bibinfo {title} {Two-dimensional
  superconductivity with strong spin-orbit interaction},\ }\href
  {https://doi.org/10.1103/PhysRevB.65.144508} {\bibfield  {journal} {\bibinfo
  {journal} {Phys. Rev. B}\ }\textbf {\bibinfo {volume} {65}},\ \bibinfo
  {pages} {144508} (\bibinfo {year} {2002})}\BibitemShut {NoStop}%
\bibitem [{\citenamefont {Fujimoto}(2005)}]{PhysRevB.72.024515}%
  \BibitemOpen
  \bibfield  {author} {\bibinfo {author} {\bibfnamefont {S.}~\bibnamefont
  {Fujimoto}},\ }\bibfield  {title} {\bibinfo {title} {Magnetoelectric effects
  in heavy-fermion superconductors without inversion symmetry},\ }\href
  {https://doi.org/10.1103/PhysRevB.72.024515} {\bibfield  {journal} {\bibinfo
  {journal} {Phys. Rev. B}\ }\textbf {\bibinfo {volume} {72}},\ \bibinfo
  {pages} {024515} (\bibinfo {year} {2005})}\BibitemShut {NoStop}%
\bibitem [{\citenamefont {Lu}\ and\ \citenamefont
  {Yip}(2008)}]{PhysRevB.77.054515}%
  \BibitemOpen
  \bibfield  {author} {\bibinfo {author} {\bibfnamefont {C.-K.}\ \bibnamefont
  {Lu}}\ and\ \bibinfo {author} {\bibfnamefont {S.}~\bibnamefont {Yip}},\
  }\bibfield  {title} {\bibinfo {title} {Signature of superconducting states in
  cubic crystal without inversion symmetry},\ }\href
  {https://doi.org/10.1103/PhysRevB.77.054515} {\bibfield  {journal} {\bibinfo
  {journal} {Phys. Rev. B}\ }\textbf {\bibinfo {volume} {77}},\ \bibinfo
  {pages} {054515} (\bibinfo {year} {2008})}\BibitemShut {NoStop}%
\bibitem [{\citenamefont {Uchihashi}(2021)}]{uchihashi2021surface}%
  \BibitemOpen
  \bibfield  {author} {\bibinfo {author} {\bibfnamefont {T.}~\bibnamefont
  {Uchihashi}},\ }\bibfield  {title} {\bibinfo {title} {Surface atomic-layer
  superconductors with rashba/zeeman-type spin-orbit coupling},\ }\href
  {https://link.springer.com/article/10.1007/s43673-021-00028-x#citeas}
  {\bibfield  {journal} {\bibinfo  {journal} {AAPPS Bulletin}\ }\textbf
  {\bibinfo {volume} {31}},\ \bibinfo {pages} {27} (\bibinfo {year}
  {2021})}\BibitemShut {NoStop}%
\bibitem [{\citenamefont {Huang}\ and\ \citenamefont
  {Hoffman}(2017)}]{doi:10.1146/annurev-conmatphys-031016-025242}%
  \BibitemOpen
  \bibfield  {author} {\bibinfo {author} {\bibfnamefont {D.}~\bibnamefont
  {Huang}}\ and\ \bibinfo {author} {\bibfnamefont {J.~E.}\ \bibnamefont
  {Hoffman}},\ }\bibfield  {title} {\bibinfo {title} {Monolayer fese on
  srtio3},\ }\href {https://doi.org/10.1146/annurev-conmatphys-031016-025242}
  {\bibfield  {journal} {\bibinfo  {journal} {Annual Review of Condensed Matter
  Physics}\ }\textbf {\bibinfo {volume} {8}},\ \bibinfo {pages} {311} (\bibinfo
  {year} {2017})}\BibitemShut {NoStop}%
\bibitem [{\citenamefont {Zakeri}\ \emph {et~al.}(2023)\citenamefont {Zakeri},
  \citenamefont {Rau}, \citenamefont {Jandke}, \citenamefont {Yang},
  \citenamefont {Wulfhekel},\ and\ \citenamefont {Berthod}}]{zakeri2023direct}%
  \BibitemOpen
  \bibfield  {author} {\bibinfo {author} {\bibfnamefont {K.}~\bibnamefont
  {Zakeri}}, \bibinfo {author} {\bibfnamefont {D.}~\bibnamefont {Rau}},
  \bibinfo {author} {\bibfnamefont {J.}~\bibnamefont {Jandke}}, \bibinfo
  {author} {\bibfnamefont {F.}~\bibnamefont {Yang}}, \bibinfo {author}
  {\bibfnamefont {W.}~\bibnamefont {Wulfhekel}},\ and\ \bibinfo {author}
  {\bibfnamefont {C.}~\bibnamefont {Berthod}},\ }\bibfield  {title} {\bibinfo
  {title} {Direct probing of a large spin--orbit coupling in the fese
  superconducting monolayer on sto},\ }\href
  {https://pubs.acs.org/doi/10.1021/acsnano.3c02876} {\bibfield  {journal}
  {\bibinfo  {journal} {ACS nano}\ } (\bibinfo {year} {2023})}\BibitemShut
  {NoStop}%
\bibitem [{\citenamefont
  {Kopnin}(2001)}]{10.1093/acprof:oso/9780198507888.001.0001}%
  \BibitemOpen
  \bibfield  {author} {\bibinfo {author} {\bibfnamefont {N.}~\bibnamefont
  {Kopnin}},\ }\href
  {https://doi.org/10.1093/acprof:oso/9780198507888.001.0001} {\emph {\bibinfo
  {title} {{Theory of Nonequilibrium Superconductivity}}}}\ (\bibinfo
  {publisher} {Oxford University Press},\ \bibinfo {year} {2001})\BibitemShut
  {NoStop}%
\end{thebibliography}%

\end{document}